\documentclass[10pt,a4paper]{article}
\usepackage[utf8]{inputenc}
\usepackage[english]{babel}
\usepackage{amsmath}
\usepackage{amsfonts}
\usepackage{amssymb}
\usepackage{graphicx}
\usepackage{geometry}
\usepackage{color}
\usepackage{slashed}
\usepackage{float}

\usepackage{hyperref}
\usepackage{ifpdf}
\definecolor{nicered}{rgb}{0.7,0.1,0.1}
\definecolor{nicegreen}{rgb}{0.1,0.5,0.1}
\hypersetup{colorlinks,citecolor= nicegreen,linkcolor= nicered}

\usepackage{appendix}

\usepackage{axodraw}

\usepackage{appendix}

\usepackage{breqn}
\begin{document}


\begin{center}

{\Large \bf Lepton polarization asymmetries in rare semi-tauonic $ b \rightarrow s $ exclusive decays at FCC-$ee$}
\vspace{5mm}

{J. F. Kamenik,$^{a,b}$ S. Monteil,$^{c}$ A. Semkiv,$^{c,d}$ L. Vale Silva$^{a}$}

\vspace{3mm}

{\it \small
$^{a}$Jo\v{z}ef Stefan Institute, Jamova 39, 1000 Ljubljana, Slovenia \\
$^{b}$Faculty of Mathematics and Physics, University of Ljubljana, Jadranska 19, 1000 Ljubljana, Slovenia \\
$^{c}$ Universit\'e Clermont Auvergne, CNRS/IN2P3, LPC, F-63000 Clermont-Ferrand, France \\
$^{d}$  Taras Shevchenko National University of Kyiv, 01601 Kyiv, Ukraine
}

\end{center}

We consider measurements of exclusive rare semi-tauonic $b$-hadron decays, mediated by the  $b \rightarrow s \tau^+ \tau^-$ transition, at a future high energy circular electron-positron collider (FCC-$ee$). We argue that the high boosts of $b$-hadrons originating from on-shell $Z$ boson decays allow for a full reconstruction of the decay kinematics in hadronic $\tau$ decay modes (up to discrete ambiguities). This, together with the potentially large statistics of $Z\to b\bar b$, opens the door for the experimental determination of $\tau$ polarizations in these rare $b$-hadron decays. In light of the current experimental situation on lepton flavor universality in rare semileptonic $B$ decays, we discuss the complementary short-distance physics information carried by the $\tau$ polarizations and suggest suitable theoretically clean observables in the form of single- and double-$\tau$ polarization asymmetries.

%
\section{Introduction}
%

Rare (semi)leptonic $b$-hadron decays allow for some of the most sensitive tests of the standard model (SM) description of flavor. Consequently they constitute powerful probes of possible flavor dynamics beyond the SM. In recent years these processes have attracted a lot of attention, in part due to several experimental measurements  defying theoretical expectations within the SM. Deviations are found in $ B \rightarrow K^{(\ast)} \mu^+ \mu^- $~\cite{Aaij:2014pli} (cf. \cite{Aaij:2016flj}) and $ B_s \rightarrow \phi \mu^+ \mu^- $~\cite{Aaij:2015esa} branching ratios, as well as in the angular analysis of $ B \rightarrow K^\ast \mu^+ \mu^- $ decays at large $ K^\ast $ recoil energies~\cite{Aaij:2015oid,Abdesselam:2016llu,Wehle:2016yoi,ATLAS:2017dlm}, see also \cite{CMS:2017ivg}. While not conclusive at present, these results may constitute first hints of new physics (NP). Interestingly enough, an anomaly is also found in the measurements of the ratios $ R^{(*)}_K = \mathcal{B}(B \rightarrow K^{(*)} \mu^+ \mu^-) / \mathcal{B}(B \rightarrow K^{(*)} e^+ e^-) $~\cite{Bifani:2017,Aaij:2014ora}, thus suggesting violations of lepton flavor universality (LFU). Such effects can only arise from physics beyond the SM. Unexpected phenomena are currently also observed in charged current mediated semileptonic $ B $ meson decays involving $ \tau $ leptons in the final state: the measurements of $ R_{D^{(\ast)}} = \mathcal{B} ({B} \rightarrow D^{(\ast)} \tau^- \bar{\nu}) / \mathcal{B} ({B} \rightarrow D^{(\ast)} \ell^- \bar{\nu}) $  ratios, where $ \ell = e, \mu $~\cite{Lees:2013uzd,Aaij:2015yra} (see also \cite{Hirose:2016wfn}), exhibit tensions with the corresponding very precise SM predictions~\cite{Kamenik:2008tj, Lattice:2015rga, Bigi:2016mdz, Na:2015kha, Fajfer:2012vx,Bernlochner:2017jka}, therefore again pointing towards possible violations of LFU and thus physics beyond the SM.

On the other hand, the rare semi-tauonic decay process $ b \rightarrow s \tau^+ \tau^- $ has not been observed so far. The present upper bound on the branching ratio $ \mathcal B (B^+ \rightarrow K^+ \tau^+ \tau^-)  <  \mathcal{O} ( 10^{-3} )$ \cite{TheBaBar:2016xwe} is expected to be improved by one or two orders of magnitude in the coming decade by the Belle II experiment. Unfortunately, this is still far above the SM-predicted rates of $ \mathcal{O} ( 10^{-7} ) $~\cite{Bouchard:2013mia} (and Section~\ref{sec:SMpredictions} below). Consequently, possible NP effects in rare semi-tauonic $ B $ meson decays are poorly constrained at  present, cf. Ref.~\cite{Grossman:1996qj, Bobeth:2011st, Alonso:2015sja}, and the situation is not expected to improve much in the near future. 

We want to argue however that a new generation of high energy particle collider experiments could allow to constrain the relevant $ (\bar{s} b) \, (\bar{\tau} \tau) $ operators much better. The NP case for the next generation collider facilities is usually built around the SM electroweak hierarchy problem, particle dark matter and heavy neutrinos (see e.g.~\cite{Aicheler:2012bya,Baer:2013cma,CEPC-SPPCStudyGroup:2015csa,Golling:2016gvc} and \cite{deSimone:2014pda,Dev:2016dja,Blondel:2014bra,Antusch:2016vyf}). However, the high involved costs and risks motivate an exhaustive set of applications. Indeed the worst-case scenario, where no new resonances are seen at the highest available collider energies,  could well materialize. In that case, the capacity of flavor processes in general, and flavor changing neutral currents in particular, to unveil short-distance physics at very high scales, potentially much above direct collider reach, would become invaluable. In the present analysis we focus on the potentialities of a Future Circular Collider, and more particularly its electron-positron collider phase (FCC-$ee$)~\cite{Golling:2016gvc}, previously known as TLEP~\cite{Gomez-Ceballos:2013zzn}. The high luminosity of the FCC-$ee$ machine complemented by an excellent vertexing system of the FCC-$ee$ detectors under consideration would allow for unique $b$-hadron rare decay studies, for instance those involving final state tau leptons. Moreover, decays with tau leptons in the final state open up novel NP search opportunities. Contrary to light leptons, polarizations of final state taus can be in principle reconstructed through their hadronic decay kinematics. This in turn allows to construct a new class of observables, with complementary sensitivities to short-distance physics, as it will become clear later in the text.

The  $ \tau $ lepton polarization observables have already been discussed in the past \cite{Hewett:1995dk, Kruger:1996cv, Geng:1998pu, Fukae:1999ww, Aliev:1999gp, Bensalem:2002ni, Cornell:2004cp, Colangelo:2006gv, Biancofiore:2014wpa} (see also \cite{Chen:2001sj,Choudhury:2002fk,Turan:2005as,Saddique:2008xj,Lu:2011jm} for other final states, and \cite{Ivanov:2017mrj} in the context of semileptonic charged currents). Here we present a proof of principle study of the viability of the complete $B\to K^{(*)} \tau^+ \tau^-$  reconstruction at the FCC-$ee$ (in Section~\ref{sec:exp}). We then focus on the precision with which polarization asymmetries can be predicted within the SM as well as their impact on NP directions singled out by the current experimental anomalies in rare semi-muonic $B$ decays (in Section~\ref{sec:th}). Our main conclusions are summarized in Section~\ref{sec:Conclusions}. Some of the lengthier and more technical derivations and expressions are relegated to the Appendices.

%
\section{The $B^0 \to K^{*0}(892) \tau^+\tau^-$ experimental reconstruction and sensitivity at high luminosity $Z$-factory}\label{sec:exp}

%
%

A possible long-term strategy for high-energy physics at colliders, after the exploitation of the LHC and its High Luminosity upgrade, considers a tunnel in the Geneva area of about 100~km circumference, which takes advantage of the present CERN accelerator complex. The Future Circular Collider (FCC) concept builds upon the successful experience and outcomes of the LEP-LHC machines. Therefore, a possible first step of the project is to fit in the tunnel a high-luminosity $e^+e^-$ collider aimed at studying comprehensively the electroweak scale with centre-of-mass energies ranging from the $Z$ pole up to beyond the $t\bar{t}$ production threshold~\cite{Gomez-Ceballos:2013zzn}. A 100 TeV proton proton collider is then considered as the ultimate goal of the project. Let us mention that an electron proton collider is also considered as an option of this project. 

%
%

\begin{figure} 
\begin{center}
\includegraphics[width=0.7\textwidth]{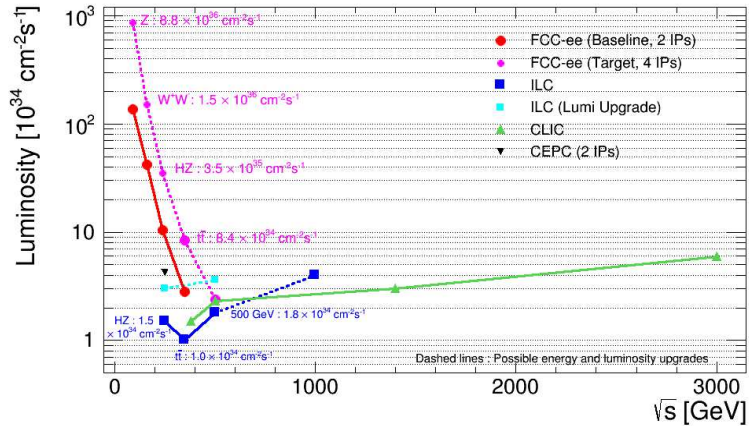} 
\caption{Comparison of $e^+e^-$ collider luminosities.}
\label{fig1}
\end{center}
\end{figure}

The goal of the high luminosity $e^+e^-$ collider is to provide collisions in the beam energy range from 40 GeV to 175 GeV. This would allow to study with unprecedented precision the four electroweak energy thresholds: 91 GeV ($Z$-pole), 160 GeV ($W$-pair production), 240 GeV (Higgs production in association with a $Z$-boson) and 350 GeV ($t \bar{t}$-pair production). In particular, the circulation of about 10000 bunches for operation at the $Z$-pole allows to envision  the production of ${\cal O}(10^{12-13})$ $Z$ decays. 
The Figure~\ref{fig1} gathers the luminosity profiles of several $e^+e^-$ collider projects and supports the above-mentioned event yields at the $Z$-pole. 

%
%

The decay $\bar B^0 \to K^{*0}(892) \tau^+ \tau^-$ is characterised by at least two neutrinos in the final state. Their presence makes the corresponding experimental search very challenging.  However, an excellent knowledge of the decay vertices, which can be obtained thanks to the multibody hadronic $\tau$ decays, may help to fully solve the kinematics of these decays. Namely, the reconstruction of the primary vertex and the decay vertex of the $B^0$ are defining the direction of the $B^0$ meson, fixing two degrees of freedom. The reconstruction of the two $\tau$ leptons' decay vertices are providing four further constraints (not all independent though). Eventually, the knowledge of the mass of the $\tau$ lepton closes the system, up to a quadratic ambiguity. FCC-$ee$ experiments are defining unique features to perform this kinematical fit: the clean leptonic machine environment allowing to place the vertex detector as close as 2~cm from the interaction point and the boost experienced by the $B^0$-meson at the $Z$-pole. 

We have studied the $\bar B^0 \to K^{*0}(892) \tau^+ \tau^-$ decay in the context of the FCC-$ee$ machine by means of Monte Carlo simulated events (signal and background) generated with a fast simulation featuring a parametric detector. The detector performance considered in the study is inspired by that obtained for the ILD vertex detector~\cite{Behnke:2013lya}.  For the sake of simplicity we have chosen  to consider the $\tau$ decays into three charged pions (mostly proceeding through the two-body process $\tau^- \to a_1^- \nu_{\tau}$ and its charge conjugate).   

\begin{figure} 
\begin{center}
\includegraphics[width=0.65\textwidth]{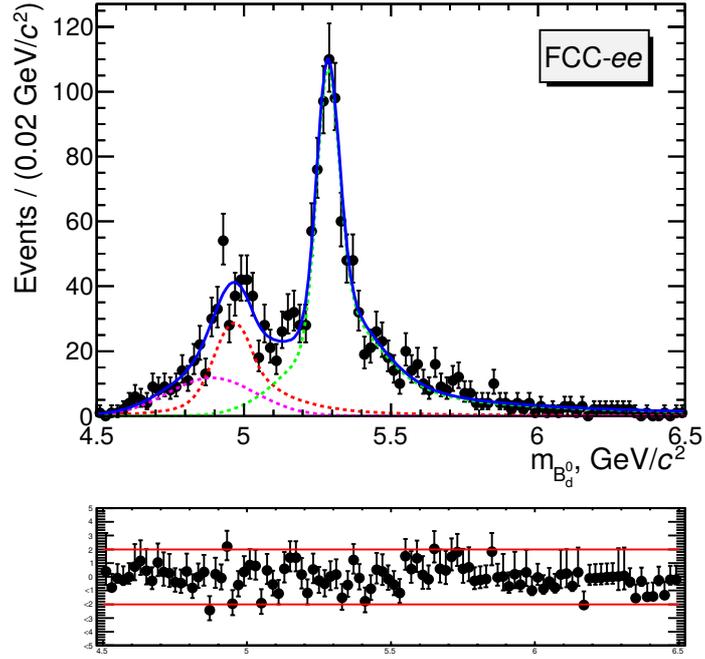} 
\caption{Invariant mass reconstruction of $\bar B^0 \to K^{*0}(892) \tau^+ \tau^- $ candidates. The $\tau$ particles are decaying into three prongs $\tau^- \to \pi^-\pi^+\pi^- \nu_{\tau}$ allowing the $\tau$ decay vertex to be reconstructed. The primary vertex ($Z$ vertex) is reconstructed from primary tracks and the secondary vertex ($\bar B^0$ vertex) is reconstructed thanks to the $K^*(892)$ daughter particles ($K^*(892) \to K^+ \pi^-$). Two dominant sources of backgrounds are included in the analysed sample, namely  $\bar B_s \to D_s^+D_s^- K^{*0}(892)$ and $\bar B^0 \to D_s^+ \bar K^{*0}(892) \tau^- \bar \nu_{\tau}$. They are modelled by the red and pink probability density functions (p.d.f.), respectively. The signal p.d.f. is displayed in green.     
}
\label{fig2}
\end{center}
\end{figure}

Figure~\ref{fig2} displays the reconstructed invariant mass distribution of signal and background events, simulated according to the branching fractions predicted in the SM,  and corresponding to $10^{13}$ $Z$-boson decays. About a thousand of events, cleanly reconstructed, can be expected, opening the way to measurements of the angular properties of the decay. To our knowledge, these  FCC-$ee$ performances are unequalled at any current or foreseeable experiment.  The reconstruction of the $\tau$ decays into three charged pions and an additional $\pi^0$ can be also used  in the partial reconstruction of the decay, doubling the expected signal yields.  The fully charged 5-body decays are relevant as well for the partial reconstruction technique, providing an additional 20\% statistics to the baseline study.   

%

%
\section{Tau polarization observables in rare semi-tauonic  $B_{(s)}$ decays: SM expectations and NP sensitivity}\label{sec:th}
%

\subsection{Effective Hamiltonian}\label{sec:effHamiltonian}

In the SM, the effective weak Hamiltonian relevant for the quark-level $ b \rightarrow s \tau^+ \tau^-$ transitions  at scales $ \mu \ll \mu_W \sim \mathcal{O} ( M_W, m_t ) $ is
\begin{eqnarray}\label{eq:SMHamiltonian}
\mathcal{H}^{\rm eff}_{\rm SM} &=& - 4 \frac{G_F}{\sqrt{2}} V^{}_{tb} V^{\ast}_{ts} \Bigg( C_1 (\mu) \, O^c_1 (\mu) + C_2 (\mu) \, O^c_2 (\mu) + \sum^{8}_{i=3} C_i (\mu) \, O_i (\mu) \\
&& \qquad\qquad\qquad + C^{\tau \tau}_9 (\mu) \, O_9 (\mu) + C^{\tau \tau}_{10} (\mu) \, O_{10} (\mu) \Bigg) +\rm  h.c. \,. \nonumber
\end{eqnarray}
In order to simplify the discussion in Section~\ref{sec:SMpredictions} we neglect doubly Cabibbo-suppressed contributions proportional to $ V^{}_{ub} V^{\ast}_{us} \sim \mathcal{O} (\lambda^2) V^{}_{tb} V^{\ast}_{ts} $. We use the standard operator basis
\begin{align}
& O^c_1 = (\bar{s}_L \gamma_\mu T^a c_L) (\bar{c}_L \gamma^\mu T^a b_L) \, , &
& O^c_2 = (\bar{s}_L \gamma_\mu c_L) (\bar{c}_L \gamma^\mu b_L) \, , \\
& O_3 = (\bar{s}_L \gamma_\mu b_L) \sum (\bar{q} \gamma^\mu q) \, , &
& O_4 = (\bar{s}_L \gamma_\mu T^a b_L) \sum (\bar{q} \gamma^\mu T^a q) \, , \nonumber\\
& O_5 = (\bar{s}_L \gamma_{\mu_1} \gamma_{\mu_2} \gamma_{\mu_3} b_L) \sum (\bar{q} \gamma^{\mu_1} \gamma^{\mu_2} \gamma^{\mu_3} q) \, , &
& O_6 = (\bar{s}_L \gamma_{\mu_1} \gamma_{\mu_2} \gamma_{\mu_3} T^a b_L) \sum (\bar{q} \gamma^{\mu_1} \gamma^{\mu_2} \gamma^{\mu_3} T^a q) \, , \nonumber\\
& O_7 = \frac{e}{16 \pi^2} m_b (\bar{s}_L \sigma^{\mu \nu} b_R) F_{\mu \nu} \, , &
& O_8 = \frac{g_s}{16 \pi^2} m_b (\bar{s}_L \sigma^{\mu \nu} T^a b_R) G^a_{\mu \nu} \, , \nonumber\\
& O_9 = \frac{e^2}{16 \pi^2} (\bar{s}_L \gamma^\mu b_L) \bar{\tau} \gamma_\mu \tau \, , &
& O_{10} = \frac{e^2}{16 \pi^2} (\bar{s}_L \gamma^\mu b_L) \bar{\tau} \gamma_\mu \gamma_5 \tau \, , \nonumber
\end{align}
where the sums run over all light quark flavors $ q $, and $ m_b $ is the $ \overline{\rm MS} $ bottom quark mass. In the following, we will drop the superscript ``$ \tau \tau $" of the Wilson coefficients.

The matching of the full SM theory at the EW scale ($\mu_W$) onto the effective Hamiltonian of Eq.~\eqref{eq:SMHamiltonian} is discussed in Ref.~\cite{Bobeth:1999mk}. For completeness, the numerical values of the Wilson coefficients calculated up to the NNLO order in $\alpha_s(\mu_W)$ and RGE evolved to the $b$-hadron mass scale $\mu_b=4.8$\,GeV are listed in Table~\ref{tab:numValues}. We use $ m_b (m_b) = 4.2 \, {\rm GeV} $, and $ \alpha_s (M_Z) = 0.1184 $ (and $ \alpha_s (\mu_b) = 0.216 $).

\begin{table}[t]
\renewcommand{\arraystretch}{1.5}
	\centering
	\begin{tabular}{|c|c|c|c|c|c|c|c|c|c|}
	\hline
	$ C_{1} $ & $ C_{2} $ & $ C_{3} $ & $ C_{4} $ & $ C_{5} $ & $ C_{6} $ & $ \tilde{C}_7 $ & $ C^{\rm eff}_{8} $ & $ C_{9} $ & $ C_{10} $ \\
	\hline
    -0.2632 & 1.0111 & -0.0055 & -0.0806 & 0.0004 & 0.0009 & -0.2923 & -0.1663 & 4.0749 & -4.3085 \\
	\hline
	\end{tabular}
	\caption{\it SM Wilson coefficients in the $ \overline{\rm MS}$ (NDR) scheme at the scale $ \mu_b = 4.8 \, {\rm GeV} $, Ref.~\cite{DescotesGenon:2011yn,Huber:2005ig,Gambino:2003zm,Gorbahn:2004my,Bobeth:2003at,Misiak:2006ab}. Above, $ \tilde{C}_7 \equiv C_7 - \frac{1}{3} ( C_3 + \frac{4}{3} C_4 + 20 C_5 + \frac{80}{3} C_6 ) $.}\label{tab:numValues}
\end{table}

Perturbative matrix element  corrections from the four-quark operators $ O_1 - O_6 $ and the magnetic dipole $ O_8 $ at the scale $\mu_b$ are absorbed into the \textit{effective} Wilson coefficients $ C^{\rm eff}_7, C^{\rm eff}_8, C^{\rm eff}_9 $ given below (cf. e.g. \cite{Bobeth:2010wg})
\begin{eqnarray}
C^{\rm eff}_7 &\equiv & C_7 - \frac{1}{3} \left[ C_3 + \frac{4}{3} C_4 + 20 C_5 + \frac{80}{3} C_6 \right] + \frac{\alpha_s}{4 \pi} \left[ (C_1 - 6 C_2) A (q^2) - C_8 F^{(7)}_8 (q^2) \right] \, , \label{eq:effWC7}\\
C^{\rm eff}_8 &\equiv & C_8 + C_3 - \frac{1}{6} C_4 + 20 C_5 - \frac{10}{3} C_6 \, , \label{eq:effWC8}\\
C^{\rm eff}_9 &\equiv & C_9 + Y (q^2) + 8 \frac{m^2_c}{q^2} \left[ \frac{4}{9} C_1 + \frac{1}{3} C_2 + 2 C_3 + 20 C_5 \right] \label{eq:effWC9}\\
&& + \frac{\alpha_s}{4 \pi} \left[ C_1 \left( B (q^2) + 4 C (q^2) \right) - 3 C_2 \left( 2 B (q^2) - C (q^2) \right) - C_8 F^{(9)}_8 (q^2) \right] \, , \nonumber
\end{eqnarray}
together with $ C^{\rm eff}_{10} \equiv C_{10} $, where $ C^{\rm eff}_8 $ is given for completeness and will not be relevant in our discussion. The explicit dependence of the (effective) Wilson coefficients and $ \alpha_s$ on the scale $ \mu_b $ have been suppressed, and only the first power in $ m_c^2 / q^2 $ is kept (where $ m_c $ is the $ \overline{\rm MS} $ charm-quark mass, $ m_c (m_c) = 1.27 \, {\rm GeV} $), consistently with \cite{Seidel:2004jh} where the charm is treated as massless. 
In Eq.~\eqref{eq:effWC9}, $Y (q^2)$ reads 
\begin{eqnarray}\label{eq:Yequation}
Y (q^2) &=& h (q^2) \left[ \frac{4}{3} C_1 + C_2 + \frac{11}{2} C_3 - \frac{2}{3} C_4 + 52 C_5 - \frac{32}{3} C_6 \right] \\
&& - \frac{1}{2} h (m_b, q^2) \left[ 7 C_3 + \frac{4}{3} C_4 + 76 C_5 + \frac{64}{3} C_6 \right] + \frac{4}{3} \left[ C_3 + \frac{16}{3} C_5 + \frac{16}{9} C_6 \right] \, , \nonumber
\end{eqnarray}
and the functions $ A, B, C $ (or equivalently $ F^{(7)}_{1,u}, F^{(7)}_{2,u}, F^{(9)}_{1,u}, F^{(9)}_{2,u} $) are found in \cite{Seidel:2004jh}, while $ F^{(7)}_8, F^{(9)}_8 $ are found in \cite{Beneke:2001at}. Finally, the charm loop function  entering the expression of $ Y (q^2) $ is
\begin{equation}
h(q^2) = \frac{8}{27} + \frac{4}{9} \left( \log \frac{\mu^2}{q^2} + i \pi \right) \, ,
\end{equation}
and
\begin{equation}
h(m_b,q^2) = \frac{4}{9} \left( \log \frac{\mu^2}{m_b^2} + \frac{2}{3} + z \right) - \frac{4}{9} (2+z) \sqrt{z-1} \arctan \frac{1}{\sqrt{z-1}} \, , \quad z = \frac{4 m_b^2}{q^2} \, .
\end{equation}

In presence of NP as hinted to by the present $b\to s \mu^+\mu^-$ measurements, the effective Wilson coefficients can receive corrections of the form
\begin{equation}
C^{\rm eff}_i \rightarrow C^{\rm eff}_i + \delta C_i \, , \quad i = 9, 10 \, .
\end{equation}
In addition, new operators can also be induced, contributing to the processes at hand at the tree level. Here we are including the chirally flipped operators
\begin{equation}
O'_9 = \frac{e^2}{16 \pi^2} (\bar{s}_R \gamma^\mu b_R) \bar{\tau} \gamma_\mu \tau \, , \quad
O'_{10} = \frac{e^2}{16 \pi^2} (\bar{s}_R \gamma^\mu b_R) \bar{\tau} \gamma_\mu \gamma_5 \tau \, ,
\end{equation}
while additional scalar and tensorial operators, not favored by the anomalies in muon data, will not be considered. We also do not consider a shift in $ \delta C_7 $ nor the operator $ O'_7 $, which are LFU-conserving (see, e.g., \cite{Paul:2016urs} for constraints on $ \delta C_7, C'_7 $). Note that in the SM the operator $ O'_7 $ is present with a suppression factor $ m_s / m_b $, giving a (higher-order) contribution that we neglect. Of course, factors of $ M_{K^{(\ast)}} / M_B $ or $ M_\phi / M_{B_s} $ are kept throughout our analysis. In summary, we consider the following effective Hamiltonian at the scale $ \mu$
\begin{eqnarray}
\mathcal{H}^{\rm eff} &=& \mathcal{H}^{\rm eff}_{\rm SM} - 4 \frac{G_F}{\sqrt{2}} V^{}_{tb} V^{\ast}_{ts} \Bigg[ \delta C_9 (\mu) \, O_9 (\mu) + \delta C_{10} (\mu) \, O_{10} (\mu) \\
&& \qquad\qquad\qquad\qquad\qquad + C'_9 (\mu) \, O'_9 (\mu) + C'_{10} (\mu) \, O'_{10} (\mu) \Bigg] + \rm h.c.\,. \nonumber
\end{eqnarray}
\noindent
In the present analysis, we do not discuss possible UV completions of $ \mathcal{H}^{\rm eff} $ but instead refer the interested reader to the existing literature on the subject~\cite{Buras:2014fpa,Alonso:2015sja,Crivellin:2017zlb,Feruglio:2017rjo}.

\subsection{$\tau$ polarization observables}\label{sec:defsObs}

\begin{figure}[t]
\centering
\SetScale{1}\SetWidth{1.2}\begin{picture}(200,220)(0,0)
	\Line(-80,10)(80,10)
	\Line(-40,200)(120,200)
	\Line(-80,10)(-40,200)
	\Line(100.5,109)(120,200)
	\DashLine(86.5,40)(100.5,109){3}
	\Line(80,10)(86.5,40)
	\Text(-20,180)[c]{\scriptsize $ \mathcal{R}'_{\tau \tau} $}
	\Line(20,40)(180,40)
	\Line(160,160)(320,160)
	\Line(20,40)(160,160)
	\Line(180,40)(320,160)
	\LongArrow(100,109)(160,109)
	\LongArrow(100,109)(40,109)
	\Vertex(100,109){3}
	\Text(110,100)[c]{\scriptsize $ B_{(s)} $}
	\Text(160,100)[c]{\scriptsize $ K^{(\ast)}, \phi $}
	\DashLine(-60,109)(100,109){10}
	\Text(70,109)[c]{$ \int \int $}
	\LongArrow(10,109)(-10,150)
	\Text(-10,160)[c]{\scriptsize $ \tau^- $}
	\Text(0,150)[c]{\scriptsize $ z'_\tau $}
	\LongArrow(10,109)(-20,90)
	\Text(-20,80)[c]{\scriptsize $ x'_\tau $}
	\LongArrow(10,109)(30,68)
	\Text(30,60)[c]{\scriptsize $ \tau^+ $}
	\Vertex(10,109){3}
	\LongArrow(-60,109)(-100,109)
	\Text(-110,109)[c]{\scriptsize $ z_\tau $}
	\LongArrow(-60,109)(-85,88)
	\Text(-85,80)[c]{\scriptsize $ x_\tau $}
	\LongArrow(-60,109)(-60,149)
	\Text(-60,155)[c]{\scriptsize $ y_\tau $}
	\Text(-80,125)[c]{\scriptsize $ \mathcal{R}_{\tau \tau} $}
	\ArrowArc(100,109)(40,40,78)
	\Text(90,135)[c]{\scriptsize $ \chi $}
	\ArrowArc(10,109)(-30,295,0)
	\Text(-5,120)[c]{\scriptsize $ \theta_\tau $}
\end{picture}
\caption{\it Relevant reference frames. The four-momentum of the $ B_{(s)} $ meson is $ k $, while the one of the final-state meson is $ p $. The four-momenta of the $ \tau^- $ and $ \tau^+ $ are $ p_{-} $ and $ p_{+} $, respectively, and $ q^2 = (p_{-} + p_{+})^2 $ is the dilepton invariant mass. The $ \mathcal{R}_{\tau \tau}, \mathcal{R}'_{\tau \tau} $ reference frames correspond to a boost (indicated by ``$ \int \int $") from the $ B_{(s)} $ meson rest-frame towards the reference frame of the $ \tau^+ \tau^- $ pair, and the reference frame $ \mathcal{R}_{\tau \tau} $ is determined from the rotation of $ \mathcal{R}'_{\tau \tau} $ (with $ \widehat{x}'_\tau, \widehat{y}'_\tau, \widehat{z}'_\tau $ constituting a right-handed frame). The longitudinal polarization of the $ \tau^- $ lepton is defined along the $ z'_\tau $ axis, while the transverse (or normal) polarization is parallel (respec., orthogonal) to the plan $ x'_\tau O z'_\tau $. A similar construction applies to the $ \tau^+ $ lepton. The angle $ \theta_\tau $ describes the orientation of the negatively charged lepton. For $ M = K^\ast, \phi $, their subsequent decay products define a plane, also indicated, with relative orientation given by $ \chi $.}\label{fig:ReferenceFrame}
\end{figure}
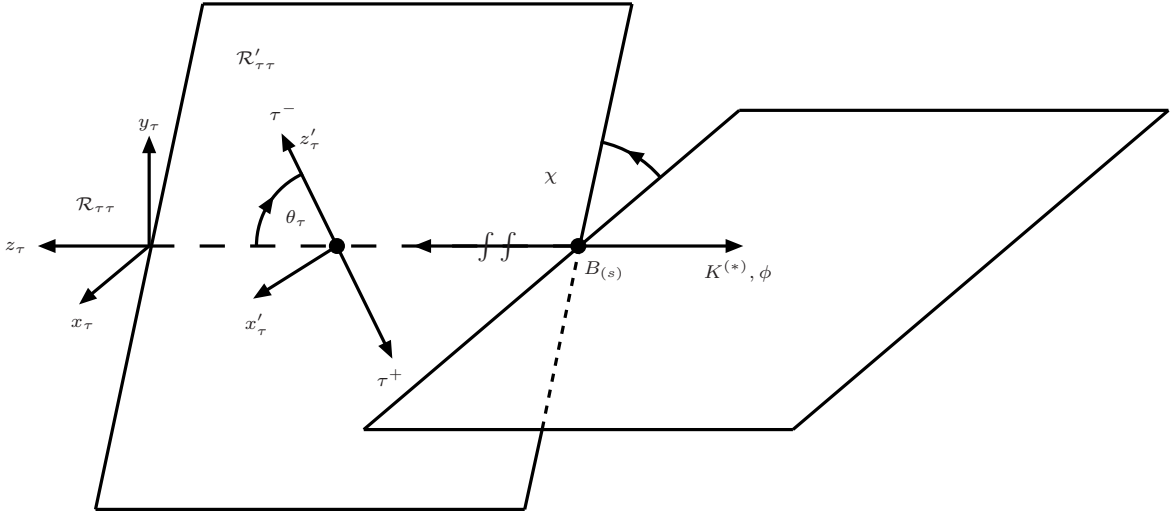

Next we introduce the (pseudo-)observables characterizing the polarizations of on-shell--produced $ \tau^\pm $ leptons \cite{Hewett:1995dk,Kruger:1996cv}. The decay kinematics of $ B_{(s)} \rightarrow M \tau^+ \tau^- $, $ M = K^{(\ast)}, \phi $, is depicted in Figure~\ref{fig:ReferenceFrame}. Note that in the following we consider stable final-state leptons and the $M$ meson, which is equivalent to working in the narrow width approximation.
The longitudinal, transverse and normal polarizations  of the $\tau^{\pm}$ can be probed using (single) polarization asymmetries $ \mathcal{P}^{\pm}_A $, for $ A = L,T,N $. They can be defined relative to a basis of space-like directions onto which we project the spinors of the $ \tau^\pm $. For definiteness, we focus first on the $ \tau^- $ where we choose
\begin{equation}\label{eq:refFramesMinus}
s^-_L = \left( 0, \frac{\overrightarrow{p}_{-}}{| \overrightarrow{p}_{-} |} \right) \, , \quad
s^-_N = \left( 0, \frac{\overrightarrow{p} \times \overrightarrow{p}_{-}}{| \overrightarrow{p} \times \overrightarrow{p}_{-} |} \right) \, , \quad
s^-_T = \left( 0, \frac{\overrightarrow{p} \times \overrightarrow{p}_{-}}{| \overrightarrow{p} \times \overrightarrow{p}_{-} |} \times \frac{\overrightarrow{p}_{-}}{| \overrightarrow{p}_{-} |} \right) \, ,
\end{equation}
with $ s^-_L, s^-_T, s^-_N $ given in the rest-frame of the $ \tau^- $ while $ \overrightarrow{p}_{-}, \overrightarrow{p} $ are the three-momenta of respectively the $ \tau^- $ lepton and the final-state $M$ meson, defined in the rest-frame of the $ \tau^- \tau^+ $ pair. The $ s^+_A $ directions (in the $\tau^+$ rest-frame) can be defined analogously with the replacement $ \overrightarrow p_{-}  \to \overrightarrow p_{+}$:
\begin{equation}\label{eq:refFrames}
s^+_L = \left( 0, \frac{\overrightarrow{p}_{+}}{| \overrightarrow{p}_{+} |} \right) \, , \quad
s^+_N = \left( 0, \frac{\overrightarrow{p} \times \overrightarrow{p}_{+}}{| \overrightarrow{p} \times \overrightarrow{p}_{+} |} \right) \, , \quad
s^+_T = \left( 0, \frac{\overrightarrow{p} \times \overrightarrow{p}_{+}}{| \overrightarrow{p} \times \overrightarrow{p}_{+} |} \times \frac{\overrightarrow{p}_{+}}{| \overrightarrow{p}_{+} |} \right) \, .
\end{equation}
Orthonormality of $s_A$ can now be used to define polarized $\tau^\pm$ spinors as $ \bar{u} (\pm e^-_A) = \bar{u} \cdot P^-_A(\pm)$ and $ v (\pm e^+_A) = P^+_A(\pm) \cdot v $, where $P^a_A (\pm) \equiv ( 1_4 \pm \gamma_5 \slashed{s}^a_A ) / 2$ such that $P^a_A (\pm)\cdot P^a_A (\pm) = P^a_A (\pm)$ and $P^a_A (\pm)\cdot P^a_A (\mp)=0$~\cite{Tsai:1971vv}. The $\tau^\pm$ polarization asymmetries (or simply polarizations) are then defined as
\begin{equation}\label{eq:singleAsymsMinus}
\mathcal{P}^{\mp}_A (q^2) = \frac{\left[ \frac{d \Gamma}{d q^2} (e^\mp_A, e^\pm_A) + \frac{d \Gamma}{d q^2} (e^\mp_A, - e^\pm_A) \right] - \left[ \frac{d \Gamma}{d q^2} (- e^\mp_A, e^\pm_A) + \frac{d \Gamma}{d q^2} (- e^\mp_A, - e^\pm_A) \right]}{\frac{d \Gamma}{d q^2} (e^\mp_A, e^\pm_A) + \frac{d \Gamma}{d q^2} (e^\mp_A, - e^\pm_A) + \frac{d \Gamma}{d q^2} (- e^\mp_A, e^\pm_A) + \frac{d \Gamma}{d q^2} (- e^\mp_A, - e^\pm_A)} \, ,
\end{equation}
for $ A = L,T,N $, after integration over the angles $ \theta_\tau $ and $ \chi_\tau $. In this definition, the indices $ \pm e^-_A $ ($ \pm e^+_A $) indicate the states that are left invariant by the projector $P^-_A(\pm)$ ($P^+_A(\pm)$). In the massless limit $ m_\tau \rightarrow 0^+ $ in the SM the longitudinal asymmetry reduces to the left-right asymmetry, while the transverse and normal asymmetries vanish. 
Also note that $\cal P_A^\pm$ only require the knowledge of one of the two polarizations. One can also consider polarization asymmetries defined for both $ \tau^\pm $ polarization states, as follows
\begin{equation}\label{eq:correlatedAsyms}
\mathcal{P}_{AB} (q^2) = \frac{\left[ \frac{d \Gamma}{d q^2} (e^-_A, e^+_B) - \frac{d \Gamma}{d q^2} (- e^-_A, e^+_B) \right] - \left[ \frac{d \Gamma}{d q^2} (e^-_A, - e^+_B) - \frac{d \Gamma}{d q^2} (- e^-_A, - e^+_B) \right]}{\frac{d \Gamma}{d q^2} (e^-_A, e^+_B) + \frac{d \Gamma}{d q^2} (e^-_A, - e^+_B) + \frac{d \Gamma}{d q^2} (- e^-_A, e^+_B) + \frac{d \Gamma}{d q^2} (- e^-_A, - e^+_B)} \, ,
\end{equation}
for $ A, B = L,T,N $. The complete analytic expressions for the polarization observables are given in Appendix~\ref{app:expAsyms}.

The first application of the tau polarization asymmetries defined above concerns the ratios of the longitudinal and transverse asymmetries. It turns out that all dependence on the Wilson coefficients cancels out in these quantities. For the decay $ B \rightarrow K \tau^+ \tau^- $ we have
\begin{equation}\label{eq:specialRelK}
\frac{\mathcal{P}^\pm_L (K) (q^2)}{\mathcal{P}^\pm_T (K) (q^2)} = \frac{f_+ (q^2)}{f_0 (q^2)} \frac{4 \, \sqrt{q^2-4 \, m_\tau^2} \, \sqrt{ \lambda (M_B^2, M_K^2, q^2) }}{3 \pi \, m_\tau \, (M_B^2 - M_K^2)} \, ,
\end{equation}
where $ \lambda (a, b, c) \equiv a^2 + b^2 + c^2 - 2 \, (a \, b + b \, c + a \, c) $. Note that this expression does not depend on the Wilson coefficients, but only on the ratio of hadronic form factors (parameterizing local matrix elements of operators entering $\mathcal H^{\rm eff}$ as defined in~Appendix~\ref{app:FFs}) $ f_+ (q^2) / f_0 (q^2) $, therefore being a clean (differential) probe of the relevant form factor determinations. Importantly, this remains true even in presence of possible NP effects of the form $ \delta C_{9, 10}, C'_{9, 10} $ (and also $ \delta C_7, C'_7 $, but not necessarily in presence of scalar and tensor semileptonic four fermion operator effects). To obtain similar observables in $ B \rightarrow K^\ast \tau^+ \tau^- $ or $ B_s \rightarrow \phi \tau^+ \tau^- $ decays, we need to project to the \textit{longitudinal} polarization of the $ K^\ast $ or $ \phi $. Then we can define

\begin{eqnarray}\label{eq:specialRelV}
&& \frac{\mathcal{P}^{\pm}_L (V_L) (q^2)}{\mathcal{P}^{\pm}_T (V_L) (q^2)} = \Bigg[ (M_{B_{(s)}} + M_{V})^2 \, (M_{B_{(s)}}^2 - M_{V}^2 - q^2) \, \frac{A_1 (q^2)}{A_0 (q^2)} - \lambda (M_{B_{(s)}}^2, M_{V}^2, q^2) \, \frac{A_2 (q^2)}{A_0 (q^2)} \Bigg] \nonumber\\
&& \qquad\qquad\qquad\qquad\qquad \times \frac{2 \sqrt{q^2 - 4 \, m_\tau^2}}{3 \pi \, M_{V} \, m_\tau \, (M_{B_{(s)}} + M_{V}) \, \sqrt{ \lambda (M_{B_{(s)}}^2, M_{V}^2, q^2) }} \, , \quad V = K^\ast, \phi \, ,
\end{eqnarray}
which is again independent of the relevant Wilson coefficients and thus defines a clean unbinned experimental constraint on the relevant form factors (or possibly indicates the presence of scalar or tensor semileptonic four fermion operators).

\subsection{Sources of uncertainty}
\label{sec:uncert}



In order to derive quantitative predictions for the observables in rare exclusive semi-tauonic $B_{(s)}$ decays, we need to evaluate the corresponding exclusive hadronic matrix elements of operators in $ \mathcal{H}^{\rm eff} $. Generically, the relevant decay amplitudes can be  written as a sum over products of the so-called helicity amplitudes $H_F^\lambda \, (\bar{u} \, \Gamma^F \, v) (\lambda)$,  $F = V, A, P, \; (S, T, T5)\,,$ times a combination of Wilson coefficients and kinematical factors. Here, $ \lambda $ is the helicity of the $ K^{(\ast)}, \phi $, or the leptonic pair, and $ \Gamma^V, \Gamma^A, \Gamma^P $ denote respectively the Lorentz structures $ \gamma^\mu, \gamma^\mu \gamma_5, \gamma_5 $ (or $ 1_4, \sigma_{\mu \nu}, \sigma_{\mu \nu} \, \gamma_5 $ for $ \Gamma^S, \Gamma^T, \Gamma^{T5} $, respec., not considered here), where the bilinear $ \bar{u} \, \gamma_5 \, v $ is derived from $ q^\mu \cdot \bar{u} \, \gamma_\mu \gamma_5 \, v $. 

An important source of theoretical uncertainty in rare semileptonic $b$-decays are the so-called charm-loop effects appearing when the $(\bar s \Gamma_1 b) (\bar c \Gamma_2 c)$ operators in $\mathcal H^{\rm eff}$ are contracted with the EM current  $ j^\mu_{e.m.} $.  Since the physical phase space of $B_{(s)} \to M \tau^+\tau^-$, $ M = K^{(\ast)}, \phi $, decays is restricted to the small hadronic recoil region, one can use the corresponding hard scale $ \sqrt{q^2} \gg E_M $ to control the size of such effects in a perturbative expansion in powers of $ \Lambda_{QCD} / m_b $ ($ \sim m_s / m_b $) and $ \alpha_s $ (a power of the latter being integrated in Eqs.~\eqref{eq:effWC7} and \eqref{eq:effWC9})~\cite{Grinstein:2004vb,Beylich:2011aq}. Of course, such a perturbative approach cannot capture the long-distance hadronic dynamics at the origin of the broad $c\bar c$ resonances such as $ \psi (3770), \psi (4040), \psi (4160), \psi (4415) $, standing above the sharp $ \psi(2S) $ peak. To take into account $ \Lambda_{QCD} / m_b $ power corrections, and corrections stemming from intermediate $c\bar c$ rescattering effects, we adopt a simplified treatment and consider an uncertainty of $ 10~\% \times C^{\rm eff}_9 $ at the level of the $ H_V^\lambda $ helicity amplitude introduced above. We also allow for arbitrary relative strong phase differences of this correction when interfering with the local matrix elements of operators in $\mathcal H^{\rm eff}$ (see also Ref.~\cite{Bobeth:2010wg,Bobeth:2011gi,Bobeth:2012vn}). In the following we refer to this set of corrections as ``Charm" (though more generally also weak annihilation topologies, for instance, are addressed in the OPE). Regarding long-distance hadronic dynamics, it is expected to give good results for inclusive observables, i.e., integrated over the full physical $ q^2 $ range~\cite{Beylich:2011aq}.
We furthermore cross-check our simplified treatment of ``Charm" long-distance effects in $q^2$ binned observables by employing a model for the charmonium and open-charm resonances~\cite{Kruger:1996cv} that is differential in $q^2$. Though the model assumes the factorization of resonant effects, known to be inaccurate \cite{Lyon:2014hpa}, the resulting uncertainty in the two approaches is similar in size. In our numerical analysis we highlight the cases where the simplified treatment of ``Charm" leads to uncertainty estimates substantially different than in the resonance model.

Another important source of uncertainties is the knowledge of the relevant hadronic form factors. The complete expressions for their parameterizations can be found in Appendix~\ref{app:FFs}. For the $ B \rightarrow K^{(\ast)} $ form factor extraction used here~\cite{Horgan:2013hoa} statistical lattice ensemble uncertainties dominate, while for $ B_s \rightarrow \phi $~\cite{Horgan:2015vla} they are an important component. Obviously, correlations between  individual parameters entering form factor parameterizations have to be taken into account (see Appendix~\ref{app:FFs} for details).

In the next Section we give numerical values for the theoretical uncertainties on asymmetries and branching ratios coming from the form factors, ``Charm" uncertainties, and a $ 2~\% $ uncertainty on the Wilson coefficients in Table~\ref{tab:numValues}, meant to include parametric uncertainties such as the one from $ \alpha_s $, as well as perturbative higher order effects, similarly to \cite{Horgan:2013pva}.\footnote{QED corrections to inclusive $b\to s \ell^+\ell^-$ decays have been evaluated in Ref.~\cite{Bobeth:2003at}. We note however that logarithmically enhanced QED corrections \cite{Huber:2015sra} proportional to $ \log (m_b^2 / m_\ell^2) $ are not expected to have the same impact in the case of taus as they have for light leptons.} Finally, we note that S-wave pollution is not expected to be a problem at large dilepton momenta~\cite{Becirevic:2012dp}.

\subsection{SM predictions}\label{sec:SMpredictions}

\begin{table}[t]
\renewcommand{\arraystretch}{1.3}
	\centering
	\begin{tabular}{|c|cccc|}
	\hline
	asyms., $ BR $ & CV & FF & ``Charm" & WC \\
	\hline
	$ \langle \mathcal{P}_L^- (K) \rangle $ & $ -0.246 $ & $ \pm 0.004 $ & $ \pm 0.006 $ & $ \pm 0.002 $ \\
	$ \langle \mathcal{P}_L^+ (K) \rangle $ & $ {+}0.246 $ & $ \pm 0.004 $ & $ \pm 0.006 $ & $ \pm 0.002 $ \\
	$ \langle \mathcal{P}_T^- (K) \rangle $ & $ -0.75 $ & $ < 0.001 $ & $ \pm 0.02 $ & $ \pm 0.006 $ \\
	$ \langle \mathcal{P}_T^+ (K) \rangle $ & $ +0.75 $ & $ < 0.001 $ & $ \pm 0.02 $ & $ \pm 0.006 $ \\
	\hline
	$ \langle \mathcal{P}^-_L (K) \rangle / \langle \mathcal{P}^-_T (K) \rangle $ & $ +0.33 $ & $ 0.005 $ & $ \pm 0.01 $ & $ < 0.001 $ \\
	\hline
	$ \langle \mathcal{P}_{LL} (K) \rangle $ & $ {+}0.30 $ & $ \pm 0.01 $ & $ \pm 0.06 $ & $ \pm 0.02 $ \\
	$ \langle \mathcal{P}_{TT} (K) \rangle $ & $ -0.68 $ & $ \pm 0.005 $ & $ \pm 0.02 $ & $ \pm 0.007 $ \\
	$ \langle \mathcal{P}_{NN} (K) \rangle $ & $ {-}0.20 $ & $ \pm 0.01 $ & $ \pm 0.09 $ & $ \pm 0.04 $ \\
	$ \langle \mathcal{P}_{LT} (K) \rangle $ & $ -0.33 $ & $ < 0.001 $ & $ \pm 0.03 $ & $ \pm 0.01 $ \\
	$ \langle \mathcal{P}_{TL} (K) \rangle $ & $ {-}0.33 $ & $ < 0.001 $ & $ \pm 0.03 $ & $ \pm 0.01 $ \\
	$ \langle \mathcal{P}_{LN} (K) \rangle $ & $ {+}0.11 $ & $ < 0.001 $ & $ \pm 0.07 $ & $ \pm 0.002 $ \\
	$ \langle \mathcal{P}_{NL} (K) \rangle $ & $ {+}0.11 $ & $ < 0.001 $ & $ \pm 0.07 $ & $ \pm 0.002 $ \\
	$ \langle \mathcal{P}_{TN} (K) \rangle $ & $ {-}0.03 $ & $ < 0.001 $ & $ \pm 0.03 $ & $ < 0.001 $ \\
	$ \langle \mathcal{P}_{NT} (K) \rangle $ & $ -0.03 $ & $ < 0.001 $ & $ \pm 0.03 $ & $ < 0.001 $ \\
	\hline
	$ \langle BR (K) \rangle \times 10^7 $ & $ 1.61 $ & $ \pm 0.07 $ & $ \pm 0.12 $ & $ \pm 0.06 \pm 0.03_{| V_{ts} |} $ \\
	\hline
	\end{tabular}
	\caption{\it  SM predictions for $ B^- \rightarrow K^- \tau^+ \tau^- $ observables. The first uncertainties correspond to the quadratic sum of the effects of the variations of the form factor uncertainties over their quoted intervals (after diagonalization to take into account the quoted correlations). The ``Charm" uncertainties correspond to corrections of $ 10~\% \times C^{\rm eff}_9 $ times an arbitrary $ \bf CP $-even phase $ {\rm e}^{ i \, \theta } $, which is allowed to vary freely. Finally, the column labeled ``WC" corresponds to the effect of variating the Wilson coefficients given in Table~\ref{tab:numValues} by $ 2~\% $, and include for the branching ratio the effect of the uncertainty on the value of $ | V_{ts} | $. The asymmetries $ \mathcal{A}_{FB} $ and $ \mathcal{P}_N^{\pm} $ vanish in this case. All uncertainty intervals are symmetrized.}\label{tab:BtoKSMvalues}
\end{table}

\begin{table}[t]
\renewcommand{\arraystretch}{1.3}
	\centering
	\begin{tabular}{|c|cccc|}
	\hline
	asyms., $ BR $ & CV & FF & ``Charm" & WC \\
	\hline
	$ \langle \mathcal{P}_L^{-} (K^\ast) \rangle $ & $ -0.56 $ & $ \pm 0.007 $ & $ \pm 0.03 $ & $ \pm 0.01 $ \\
	$ \langle \mathcal{P}_L^{+} (K^\ast) \rangle $ & $ {+}0.56 $ & $ \pm 0.007 $ & $ \pm 0.03 $ & $ \pm 0.01 $ \\
	$ \langle \mathcal{P}_T^{-} (K^\ast) \rangle $ & $ -0.53 $ & $ \pm 0.02 $ & $ \pm 0.04 $ & $ \pm 0.002 $ \\
	$ \langle \mathcal{P}_T^{+} (K^\ast) \rangle $ & $ +0.02 $ & $ \pm 0.04 $ & $ \pm 0.1 $ & $ \pm 0.01 $ \\
	$ \langle \mathcal{P}_N^{-} (K^\ast) \rangle $ & $ -0.02 $ & $ \pm 0.001 $ & $ \pm 0.01 $ & $ < 0.001 $ \\
	$ \langle \mathcal{P}_N^{+} (K^\ast) \rangle $ & $ {-}0.02 $ & $ \pm 0.001 $ & $ \pm 0.01 $ & $ < 0.001 $ \\
	\hline
	$ \langle \mathcal{P}^{-}_L \rangle / \langle \mathcal{P}^{-}_T \rangle \, (K^\ast_{\rm long.}) $ & $ +0.68 $ & $ \pm 0.03 $ & $ \pm 0.01 $ & $ < 0.001 $ \\
	\hline
	$ [ \langle \mathcal{P}^-_L \rangle^2 + \langle \mathcal{P}^-_T \rangle^2 ]^{1/2} \, (K^\ast) $ & $ + 0.77 $ & $ \pm 0.01 $ & $ \pm 0.02 $ & $ \pm 0.008 $ \\
	\hline
	$ \langle \mathcal{P}_{LL} (K^\ast) \rangle $ & $ {-}0.35 $ & $ \pm 0.02 $ & $ \pm 0.02 $ & $ \pm 0.007 $ \\
	$ \langle \mathcal{P}_{TT} (K^\ast) \rangle $ & $ +0.05 $ & $ \pm 0.03 $ & $ \pm 0.09 $ & $ \pm 0.01 $ \\
	$ \langle \mathcal{P}_{NN} (K^\ast) \rangle $ & $ {+}0.09 $ & $ \pm 0.02 $ & $ \pm 0.08 $ & $ \pm 0.01 $ \\
	$ \langle \mathcal{P}_{LT} (K^\ast) \rangle $ & $ -0.001 $ & $ \pm 0.02 $ & $ \pm 0.03 $ & $ \pm 0.007 $ \\
	$ \langle \mathcal{P}_{TL} (K^\ast) \rangle $ & $ {-}0.28 $ & $ \pm 0.009 $ & $ \pm 0.03 $ & $ \pm 0.01 $ \\
	$ \langle \mathcal{P}_{LN} (K^\ast) \rangle $ & $ {+}0.05 $ & $ \pm 0.002 $ & $ \pm 0.02 $ & $ \pm 0.001 $ \\
	$ \langle \mathcal{P}_{NL} (K^\ast) \rangle $ & $ {+}0.05 $ & $ \pm 0.002 $ & $ \pm 0.02 $ & $ \pm 0.001 $ \\
	$ \langle \mathcal{P}_{TN} (K^\ast) \rangle $ & $ {-}0.002 $ & $ \pm 0.002 $ & $ \pm 0.03 $ & $ < 0.001 $ \\
	$ \langle \mathcal{P}_{NT} (K^\ast) \rangle $ & $ -0.002 $ & $ \pm 0.002 $ & $ \pm 0.03 $ & $ < 0.001 $ \\
	\hline
	$ \langle \mathcal{A}_{FB} (K^\ast) \rangle $ & $ +0.20 $ & $ \pm 0.01 $ & $ \pm 0.03 $ & $ \pm 0.004 $ \\
	\hline
	$ \langle BR (K^\ast) \rangle \times 10^7 $ & $ 1.30 $ & $ \pm 0.09 $ & $ \pm 0.22 $ & $ \pm 0.07 \pm 0.03_{| V_{ts} |} $ \\
	\hline
	\end{tabular}
	\caption{\it SM predictions for the $ \bar{B} \rightarrow \bar{K}^\ast \tau^+ \tau^- $ observables. The ``Charm" uncertainties correspond to independent corrections of $ 10~\% \times C^{\rm eff}_9 $ times arbitrary phases $ {\rm e}^{ i \, \theta_\lambda } $, $ \lambda = 0, \pm 1 $, which are allowed to vary freely and independently. In this table, $ K^\ast_{\rm long.} $ denotes the longitudinal polarization of the $ K^\ast $. See Table~\ref{tab:BtoKSMvalues} for more comments.}\label{tab:BtoKstarSMvalues}
\end{table}

In this Section we give the SM predictions for the considered observables and discuss their theoretical uncertainties. 
Some of the numerical inputs used in the calculation are given below \cite{CKMfitter,Olive:2016xmw}\footnote{Though the extraction of the CKM matrix elements is made under the hypothesis of LFU, the observables involved in such extraction do not involve the LFUV considered here.}

\begin{eqnarray}
&& | V_{tb} | = 0.999119 \; (+24, -12) \, , \quad | V_{ts} | = 0.04108 \; (+30, -57) \, , \nonumber\\
&& M_B = 5.28 \; {\rm GeV} \, , \quad M_{B_s} = 5.366 \; {\rm GeV} \, , \quad m_\tau = 1.777 \, {\rm GeV} \,,\nonumber \\
&& \tau_B = 1.520 \times 10^{-12} \; {\rm s} \, , \quad \tau_{B_s} = 1.510 \times 10^{-12} \; {\rm s} \, ,\nonumber \\
&& M_K = 0.50 \; {\rm GeV} \, , \quad M_{K^\ast} = 0.892 \; {\rm GeV} \, , \quad M_\phi = 1.020 \; {\rm GeV} \, ,
\end{eqnarray}
while the value of the hyperfine constant is $ \alpha_{EM} (\mu_b) = 1 / 133 $. The difference in the masses of the charged and the neutral $ B $ and $ K $ mesons can be neglected, and then we do not generally differentiate their charge or flavor.

We define the binned total rate, polarization observables, Eqs.~\eqref{eq:singleAsymsMinus}-\eqref{eq:correlatedAsyms}, and the FB asymmetry, Eq.~\eqref{eq:FBasym} below, via $ q^2 $ integration of the corresponding differential partial rates. For instance,

\begin{eqnarray}
\langle \mathcal{P}^{-}_A \rangle &=& \frac{1}{\langle \Gamma \rangle} \int^{q^2_{\rm max}}_{q^2_{\rm min}} dq^2 \Bigg\{ \left[ \frac{d \Gamma}{d q^2} (e^-_A, e^+_A) + \frac{d \Gamma}{d q^2} (e^-_A, - e^+_A) \right] \\
&& \qquad - \left[ \frac{d \Gamma}{d q^2} (- e^-_A, e^+_A) + \frac{d \Gamma}{d q^2} (- e^-_A, - e^+_A) \right] \Bigg\} \, , \quad A = L,T,N \, , \nonumber
\end{eqnarray}
where $ \langle \Gamma \rangle \equiv \int^{q^2_{\rm max}}_{q^2_{\rm min}} dq^2 d \Gamma / d q^2 $. Similarly, $ \langle BR \rangle =  \langle \Gamma \rangle \tau_{B_{(s)}}$. Here we consider a single bin within $ q^2_{\rm min} = 14.18 \; {\rm GeV}^2 $, in order to avoid the sharp $\psi(2S)$ resonance at $ \sqrt {q^2} = 3.686 $~GeV\,, and the zero hadronic recoil kinematical endpoint $ q^2_{\rm max} = (M_B - M_{K^{(\ast)}})^2 $ or $ (M_{B_s} - M_\phi)^2 $. Therefore, the range $ [ q^2_{\rm min}, q^2_{\rm max} ] $ includes the broad resonances $ \psi (3770) $, $ \psi (4040) $, $ \psi (4160) $, and possibly $ \psi (4415) $. As discussed in Ref.~\cite{Beylich:2011aq}, the integration over $ q^2 $ damps resonant effects up to some extent so that they are captured by the effective ``Charm" contribution introduced in the previous Section. A further reason to discuss only the binned observables is that, in any case we expect at first to measure the integrated polarization asymmetries, prior to the measurement of their differential distributions.

Our numerical SM predictions are given in Tables~\ref{tab:BtoKSMvalues}, \ref{tab:BtoKstarSMvalues} and \ref{tab:BtophiSMvalues}. We note that the central value of the $ B \rightarrow K \tau^+ \tau^- $ branching ratio  is in broad agreement ($ 12~\% $ difference) with Ref.~\cite{Bouchard:2013mia}. The difference may be at least partly traced back to different sets of form factors and is roughly consistent with stated form factor uncertainties. We note that the SM values of polarization observables exhibit a distinct hierarchy with $ \langle \mathcal{P}_T^\pm (K) \rangle, \langle \mathcal{P}_{TT} (K) \rangle $ (or $ \langle \mathcal{P}_L^{\pm} (V) \rangle, \langle \mathcal{P}_T^{-} (V) \rangle $) being the largest asymmetries in the case of $ B \rightarrow K \tau^+ \tau^- $ (respec., $ B \rightarrow K^\ast \tau^+ \tau^- $ and $ B_s \rightarrow \phi \tau^+ \tau^- $), and also the cleanest ones. For all decay channels, the uncertainties are dominated by ``Charm" effects. Moreover, the form factor uncertainties are highly suppressed in the case of $ B \rightarrow K \tau^+ \tau^- $. We also point out that there are a few numerical coincidences in these tables that are purely due to the choice of $ q^2_{\rm min} $, for instance $ | \langle \mathcal{P}_L^{\pm} (V) \rangle | \simeq | \langle \mathcal{P}_T^{-} (V) \rangle | $, or $ | \langle \mathcal{P}_L^\pm (K) \rangle | \simeq 1/4 $ while $ | \langle \mathcal{P}_T^\pm (K) \rangle | \simeq 3/4 $. 

A few more comments are in order concerning the numerical values in Tables~\ref{tab:BtoKSMvalues}-\ref{tab:BtophiSMvalues}. First, the expressions of some quantities are identical up to a sign, e.g., $ | \mathcal{P}_T^- (K) | = | \mathcal{P}_T^+ (K) | $, $ | \mathcal{P}_{LT} (K) | = | \mathcal{P}_{TL} (K) | $, etc., which can be understood in terms of discrete $ \bf C $, $ \bf P $ and $ \bf T $ symmetry transformations.  Note that the signs of the different asymmetries on the other hand depend on the convention we use for the projection vectors Eqs.~\eqref{eq:refFramesMinus}-\eqref{eq:refFrames}.
For the $ K^\ast $ and $ \phi $, their transverse polarizations are at the origin of the differences $ | \mathcal{P}_T^{-} (V) | \neq | \mathcal{P}_T^{+} (V) | $ and $ | \mathcal{P}_{LT} (V) | \neq | \mathcal{P}_{TL} (V) | $. For the latter, the numerical values of $ \langle \mathcal{P}_T^+ (V) \rangle $ and $ \langle \mathcal{P}_{LT} (V) \rangle $ are suppressed due in part to the approximate pure left-handed lepton currents we have for $ \mathcal{H}^{\rm eff}_{\rm SM}  $, i.e., $ C^{\rm eff}_9 \simeq - C^{\rm eff}_{10} $. On the other hand, the single normal asymmetry $ \mathcal{P}^\pm_N $, and the correlated asymmetries $ \mathcal{P}_{LN}, \mathcal{P}_{NL}, \mathcal{P}_{TN}, \mathcal{P}_{NT} $, depend on the imaginary phases of $ C^{\rm eff}_7 $ and $ C^{\rm eff}_9 $ and are therefore greatly suppressed in the SM. Instead, $ \mathcal{P}_{NN} $ does not vanish in the absence of complex phases. 

\begin{table}[t]
\renewcommand{\arraystretch}{1.3}
	\centering
	\begin{tabular}{|c|cccc|}
	\hline
	asyms., $ BR $ & CV & FF & ``Charm" & WC \\
	\hline
	$ \langle \mathcal{P}_L^{-} (\phi) \rangle $ & $ -0.55 $ & $ \pm 0.006 $ & $ \pm 0.03 $ & $ \pm 0.008 $ \\
	$ \langle \mathcal{P}_L^{+} (\phi) \rangle $ & $ {+}0.55 $ & $ \pm 0.006 $ & $ \pm 0.03 $ & $ \pm 0.008 $ \\
	$ \langle \mathcal{P}_T^{-} (\phi) \rangle $ & $ -0.51 $ & $ \pm 0.02 $ & $ \pm 0.04 $ & $ < 0.001 $ \\
	$ \langle \mathcal{P}_T^{+} (\phi) \rangle $ & $ +0.07 $ & $ \pm 0.04 $ & $ \pm 0.1 $ & $ \pm 0.009 $ \\
	$ \langle \mathcal{P}_N^{-} (\phi) \rangle $ & $ -0.02 $ & $ \pm 0.001 $ & $ \pm 0.01 $ & $ < 0.001 $ \\
	$ \langle \mathcal{P}_N^{+} (\phi) \rangle $ & $ {-}0.02 $ & $ \pm 0.001 $ & $ \pm 0.01 $ & $ < 0.001 $ \\
	\hline
	$ \langle \mathcal{P}^{-}_L \rangle / \langle \mathcal{P}^{-}_T \rangle \, (\phi_{\rm long.}) $ & $ +0.68 $ & $ \pm 0.01 $ & $ \pm 0.01 $ & $ < 0.001 $ \\
	\hline
	$ [ \langle \mathcal{P}^-_L \rangle^2 + \langle \mathcal{P}^-_T \rangle^2 ]^{1/2} \, (\phi) $ & $ + 0.76 $ & $ \pm 0.006 $ & $ \pm 0.03 $ & $ \pm 0.009 $ \\
	\hline
	$ \langle \mathcal{P}_{LL} (\phi) \rangle $ & $ {-}0.34 $ & $ \pm 0.01 $ & $ \pm 0.02 $ & $ \pm 0.005 $ \\
	$ \langle \mathcal{P}_{TT} (\phi) \rangle $ & $ +0.04 $ & $ \pm 0.03 $ & $ \pm 0.09 $ & $ \pm 0.009 $ \\
	$ \langle \mathcal{P}_{NN} (\phi) \rangle $ & $ {+}0.11 $ & $ \pm 0.02 $ & $ \pm 0.08 $ & $ \pm 0.01 $ \\
	$ \langle \mathcal{P}_{LT} (\phi) \rangle $ & $ -0.02 $ & $ \pm 0.02 $ & $ \pm 0.03 $ & $ \pm 0.005 $ \\
	$ \langle \mathcal{P}_{TL} (\phi) \rangle $ & $ {-}0.27 $ & $ \pm 0.01 $ & $ \pm 0.03 $ & $ \pm 0.008 $ \\
	$ \langle \mathcal{P}_{LN} (\phi) \rangle $ & $ {+}0.05 $ & $ \pm 0.002 $ & $ \pm 0.02 $ & $ \pm 0.001 $ \\
	$ \langle \mathcal{P}_{NL} (\phi) \rangle $ & $ {+}0.05 $ & $ \pm 0.002 $ & $ \pm 0.02 $ & $ \pm 0.001 $ \\
	$ \langle \mathcal{P}_{TN} (\phi) \rangle $ & $ {-}0.004 $ & $ \pm 0.002 $ & $ \pm 0.03 $ & $ < 0.001 $ \\
	$ \langle \mathcal{P}_{NT} (\phi) \rangle $ & $ -0.004 $ & $ \pm 0.002 $ & $ \pm 0.03 $ & $ < 0.001 $ \\
	\hline
	$ \langle \mathcal{A}_{FB} (\phi) \rangle $ & $ +0.18 $ & $ \pm 0.01 $ & $ \pm 0.03 $ & $ \pm 0.002 $ \\
	\hline
	$ \langle BR (\phi) \rangle \times 10^7 $ & $ 1.24 $ & $ \pm 0.09 $ & $ \pm 0.21 $ & $ \pm 0.05 \pm 0.03_{| V_{ts} |} $ \\
	\hline
	\end{tabular}
	\caption{\it SM predictions for $ \bar{B}_s \rightarrow \phi \tau^+ \tau^- $ observables. See Tables~\ref{tab:BtoKSMvalues} and \ref{tab:BtoKstarSMvalues} for more comments.}\label{tab:BtophiSMvalues}
\end{table}

{As advocated in Section~\ref{sec:defsObs}, the ratio of the unbinned longitudinal and transverse asymmetries is insensitive to the effective Wilson coefficients, which absorb the effects of charm loops. In our implementation of the approach of Ref.~\cite{Beylich:2011aq}, with a correction absorbed into $ C^{\rm eff}_9 $ independent on $ q^2 $, this property is retained to a very good extent even when considering the experimentally accessible ratio of asymmetries integrated over finite $q^2$ bins, leading to a ``Charm" uncertainty smaller than $ \pm 0.001 $ in all cases. However, since in the binned ratio the cancellation of the (effective) Wilson coefficients is due to their small variation over the $q^2$ bin,  the cancellation breaks down when employing a model of  charmonium and open-charm resonances that is differential in $q^2$~\cite{Kruger:1996cv}. Employing it here, and only here, we give an estimate of the corresponding uncertainty in the case of the ratio of longitudinal and transverse polarizations, shown in Tables~\ref{tab:BtoKSMvalues}, \ref{tab:BtoKstarSMvalues} and \ref{tab:BtophiSMvalues}.}

For $ B \rightarrow K^\ast $ and $ B_s \rightarrow \phi $ decays we also give the values for the polarization vector modulus, i.e., $ [(\mathcal{P}^-_L)^2 + (\mathcal{P}^-_T)^2]^{1/2} $. In this observable there is a somewhat better control of the ``Charm" related uncertainty due to a partial cancellation of the term sensitive to the $ \bf CP $-even phase $ \exp \{ i \; \theta_{-} \} $, which is allowed to vary freely. {Still, the resulting uncertainty is of the same order as the quoted ``Charm" uncertainties for the individual longitudinal and transverse asymmetries and consistent with the estimate within the charm resonance model~\cite{Kruger:1996cv}.}

Finally, touching upon experimental aspects, note that the processes $ B^\pm \rightarrow K^\pm \tau^+ \tau^- $, and $ B^0 \rightarrow K^{\ast 0} \tau^+ \tau^- $ or $ \bar{B}^0 \rightarrow \bar{K}^{\ast 0} \tau^+ \tau^- $ are self-tagging, while the final state with a $ \phi $ meson requires tagging to determine the flavor of the parent meson (bottom or anti-bottom). In this context we note that the expressions of the asymmetries $ \mathcal{P}^\pm_L, \mathcal{P}^\pm_T, \mathcal{P}_{LN}, \mathcal{P}_{NL}, \mathcal{P}_{TN}, \mathcal{P}_{NT} $ are $ \bf P $-odd, while the expressions of the asymmetries $ \mathcal{P}^\pm_N, \mathcal{P}_{LT}, \mathcal{P}_{TL}, \mathcal{P}_{LL}, \mathcal{P}_{TT}, \mathcal{P}_{NN} $ are $ \bf P $-even. In absence of $ \bf CP $ violation then, which is the case in the SM when terms proportional to $ V^{}_{ub} V^{\ast}_{us} $ are neglected, the former set is $ \bf C $-odd while the latter is $ \bf C $-even. Therefore, the former set vanishes in the untagged sample, while the latter does not. 

\subsection{Beyond SM effects}
\label{sec:BSM}

\begin{figure}[t]
\centering
	\includegraphics[scale=0.55]{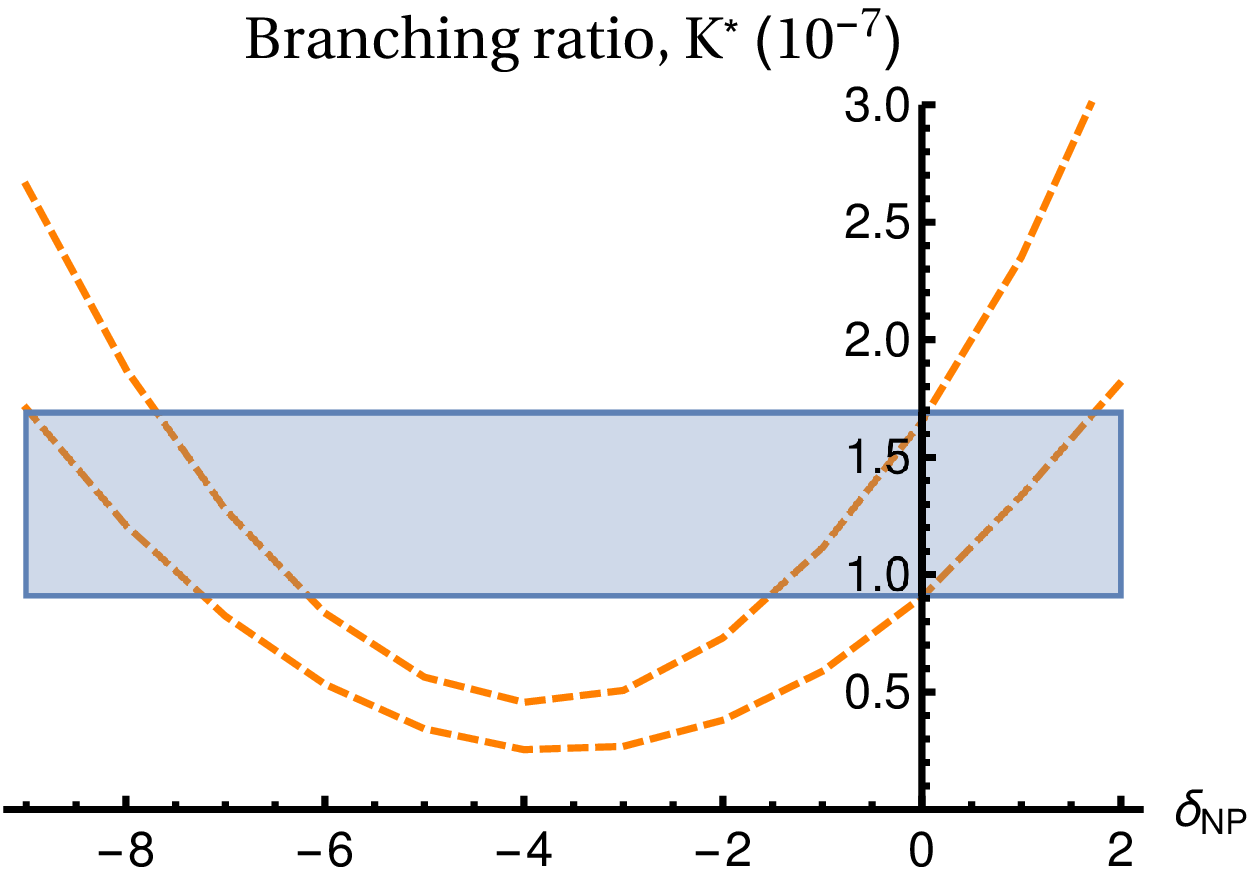}
	\hspace{3mm}\includegraphics[scale=0.55]{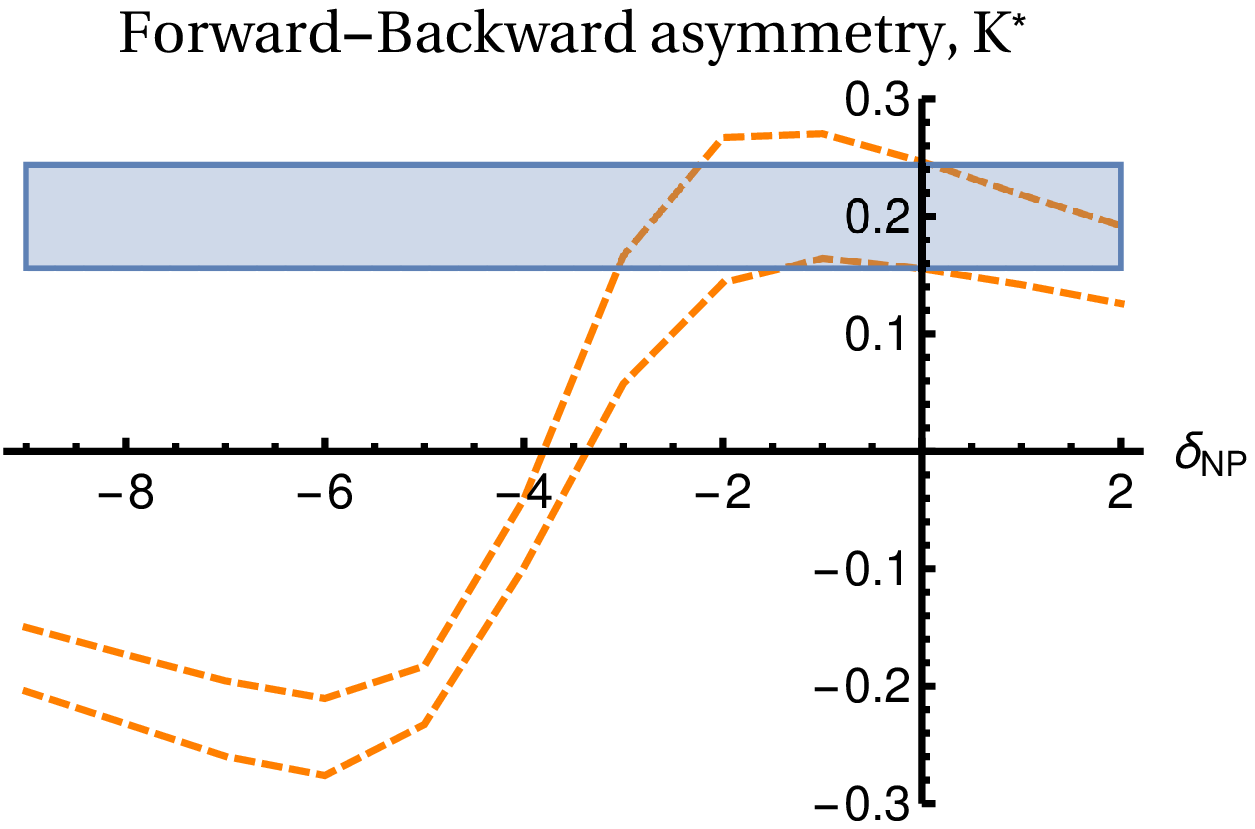} \\
	\vspace{5mm}
	\includegraphics[scale=0.55]{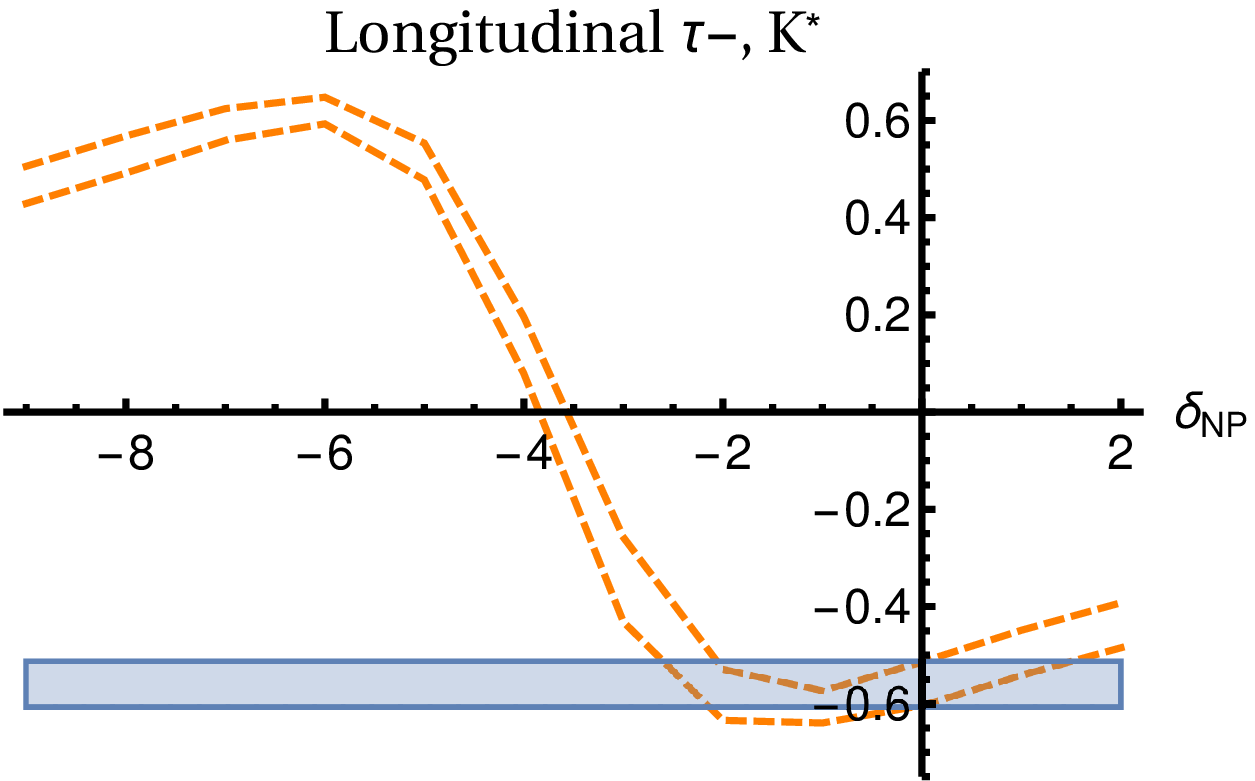}
	\hspace{3mm}\includegraphics[scale=0.55]{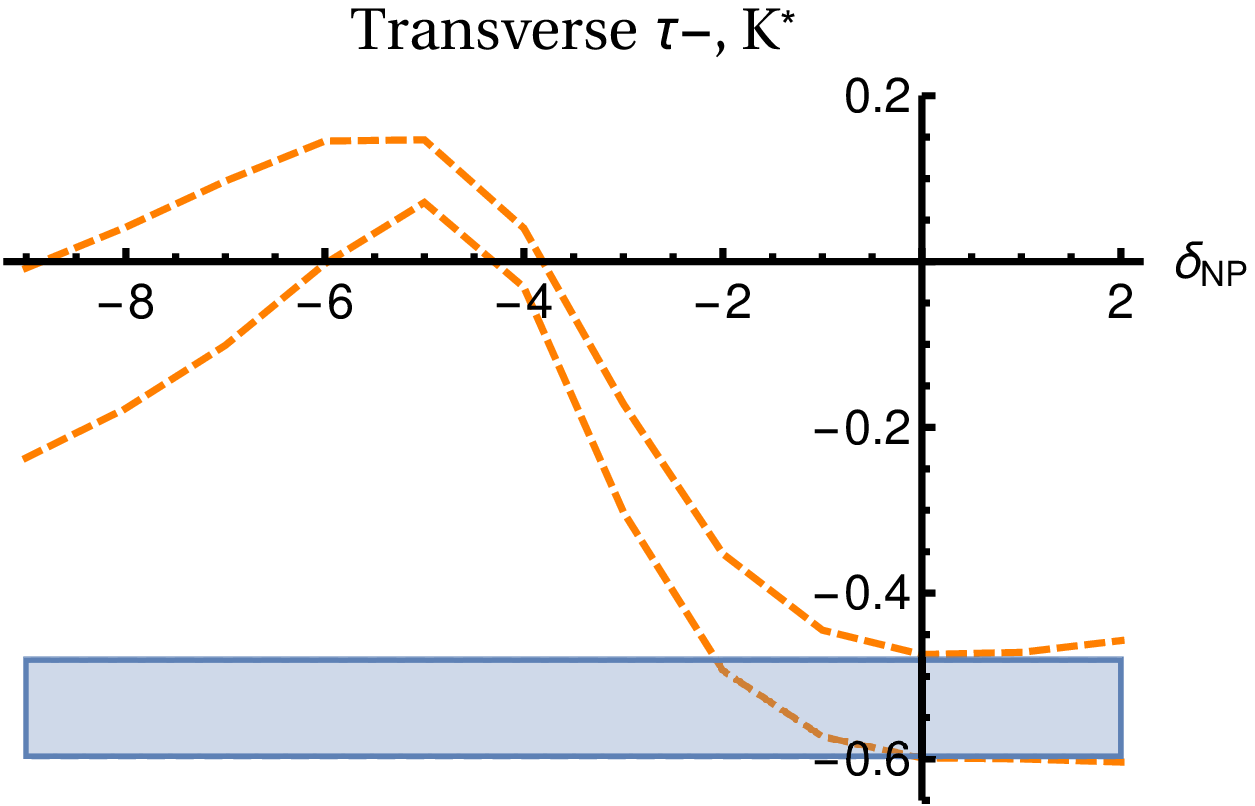}
	\caption{\it New physics effects in the form of a real $ \delta C_9 \equiv \delta_{\rm NP} $, whose values are given in the horizontal axes. The vertical axes give the values of the branching ratio, the FB asymmetry and the longitudinal and transverse polarizations of the $ \tau^- $, in the SM (solid, filled blue, independent on the value of $ \delta_{\rm NP} $) and in the NP under consideration (dashed orange). The blue band corresponds to the errors seen in Table~\ref{tab:BtoKstarSMvalues}, which are recalculated for NP. See Figure~\ref{fig:BtoKNPcase19} for plots concerning $ B \rightarrow K \tau^+ \tau^- $.}\label{fig:BtoKstarNPcase19}
\end{figure}

\begin{figure}[t]
\centering
	\includegraphics[scale=0.55]{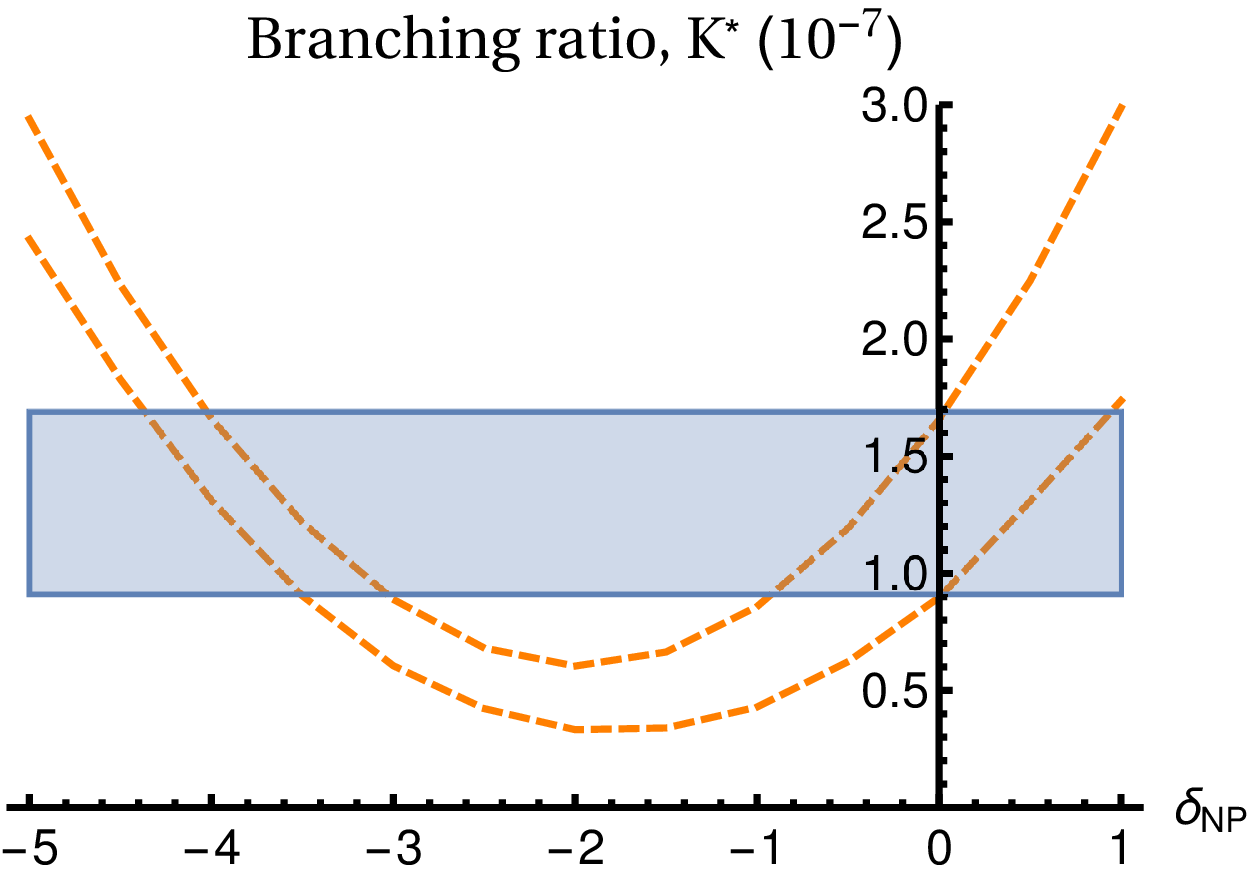}
	\hspace{3mm}\includegraphics[scale=0.55]{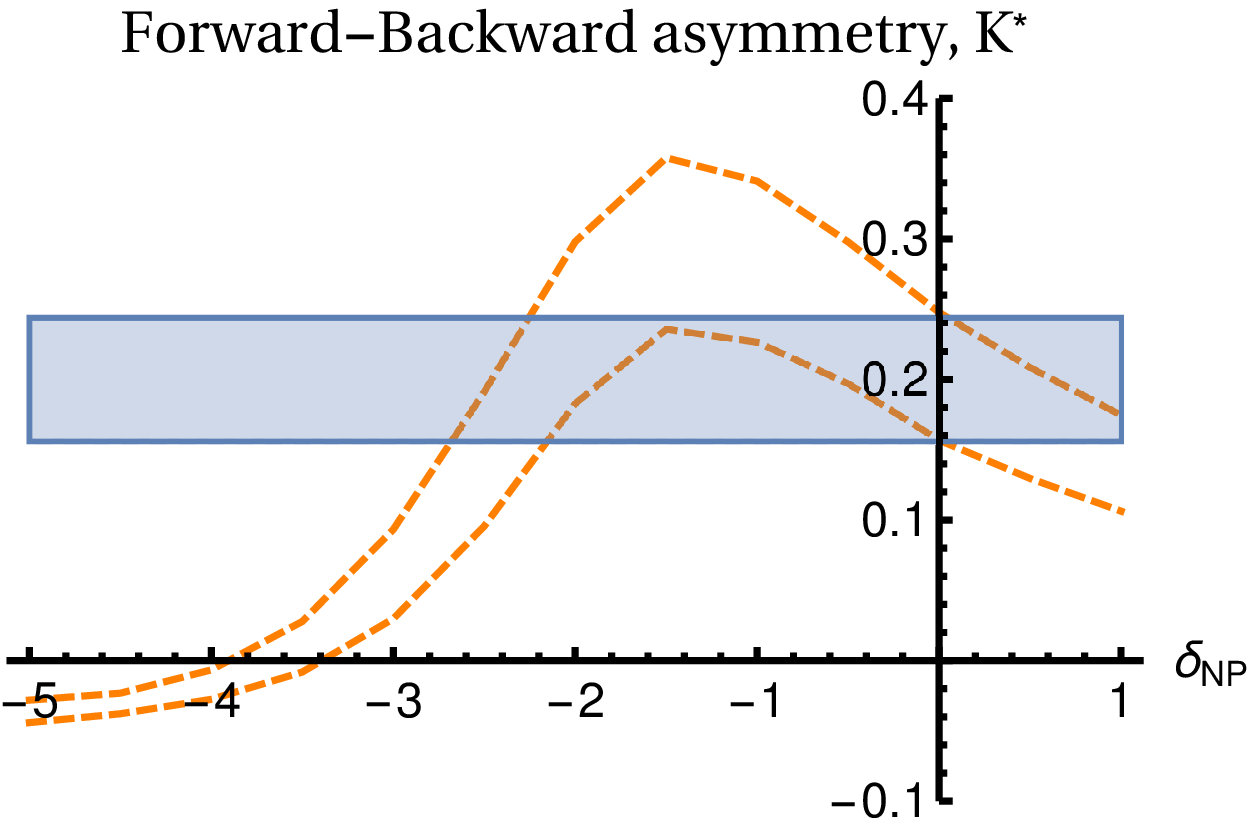} \\
	\vspace{5mm}
	\includegraphics[scale=0.55]{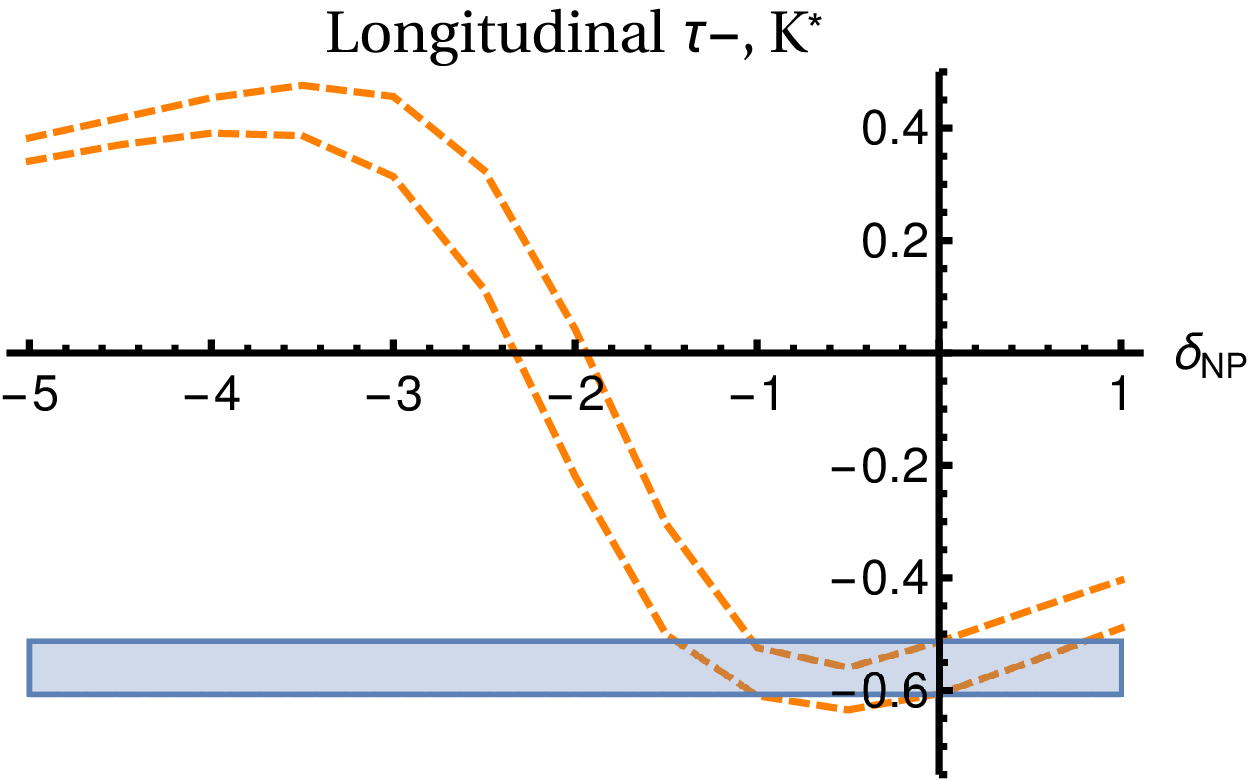}
	\hspace{3mm}\includegraphics[scale=0.55]{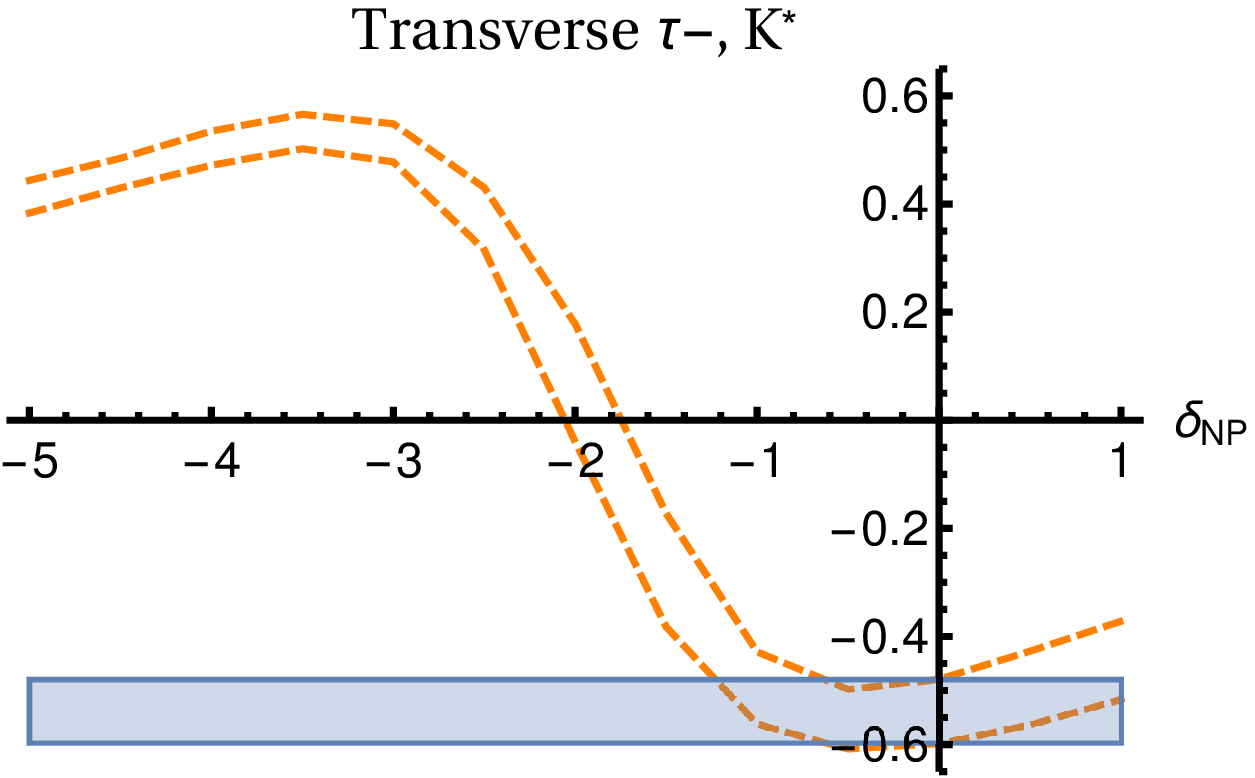}
	\caption{\it New physics effects in the form of a real $ \delta C_9 = - C'_9 \equiv \delta_{\rm NP} $, whose values are given in the horizontal axes. See Figure~\ref{fig:BtoKstarNPcase19} for more comments on the reading of the graphics.}\label{fig:BtoKstarNPcase9}
\end{figure}

\begin{figure}[t]
\centering
	\includegraphics[scale=0.55]{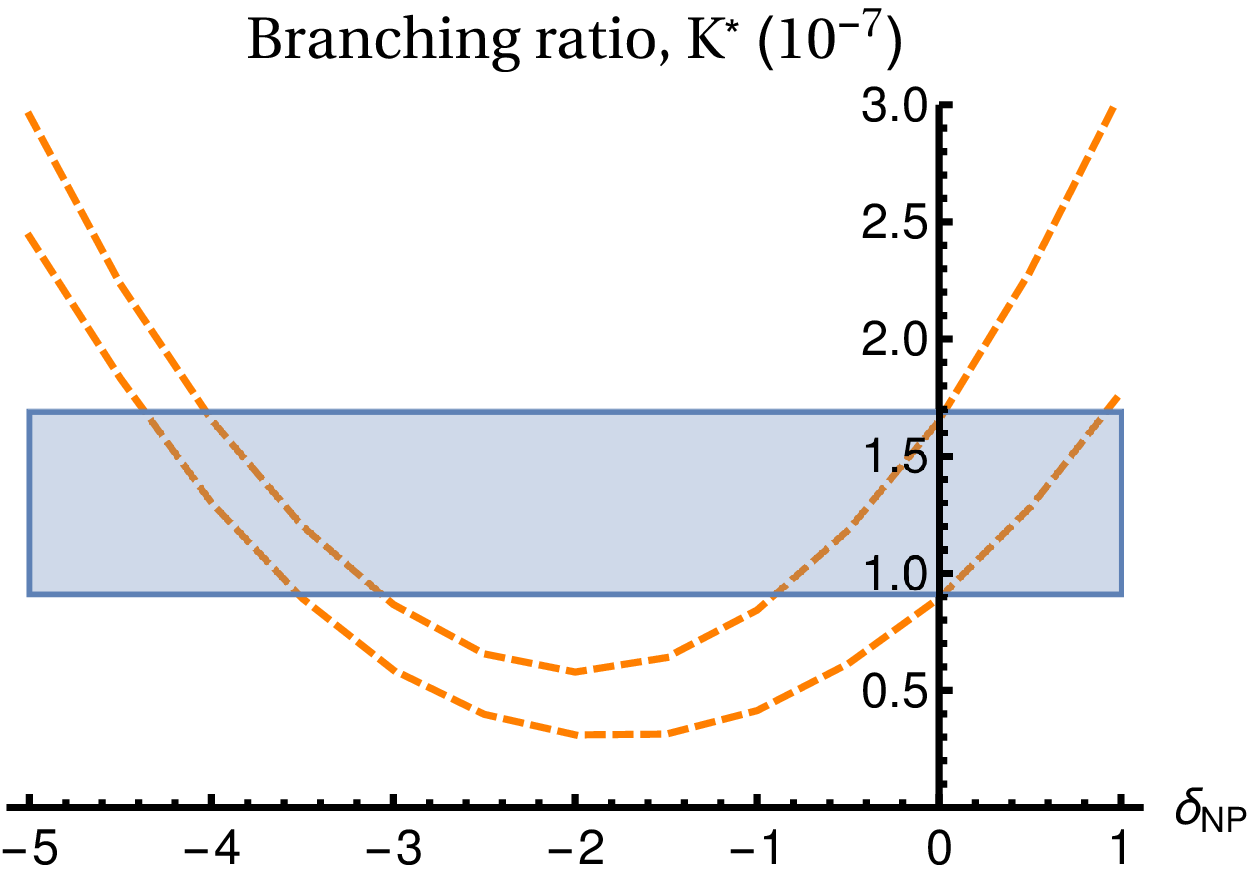}
	\hspace{3mm}\includegraphics[scale=0.55]{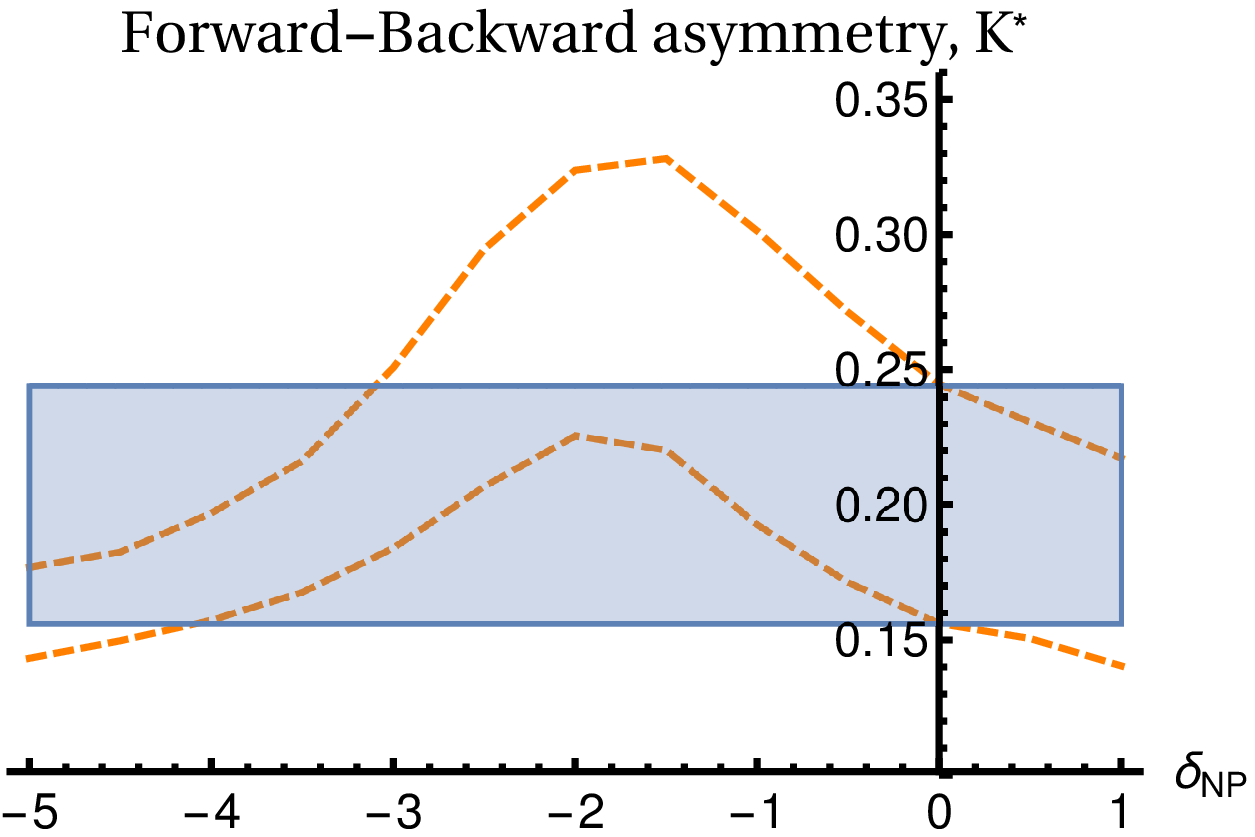} \\
	\vspace{5mm}
	\includegraphics[scale=0.55]{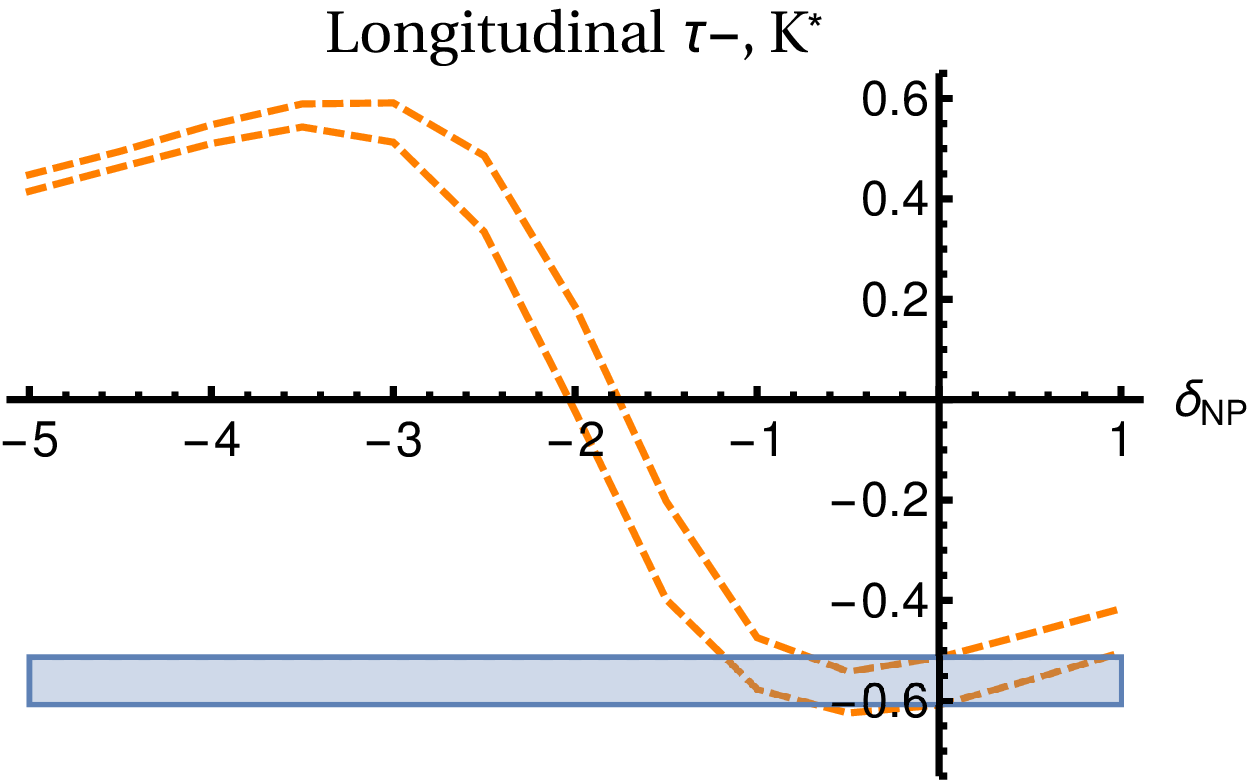}
	\hspace{3mm}\includegraphics[scale=0.55]{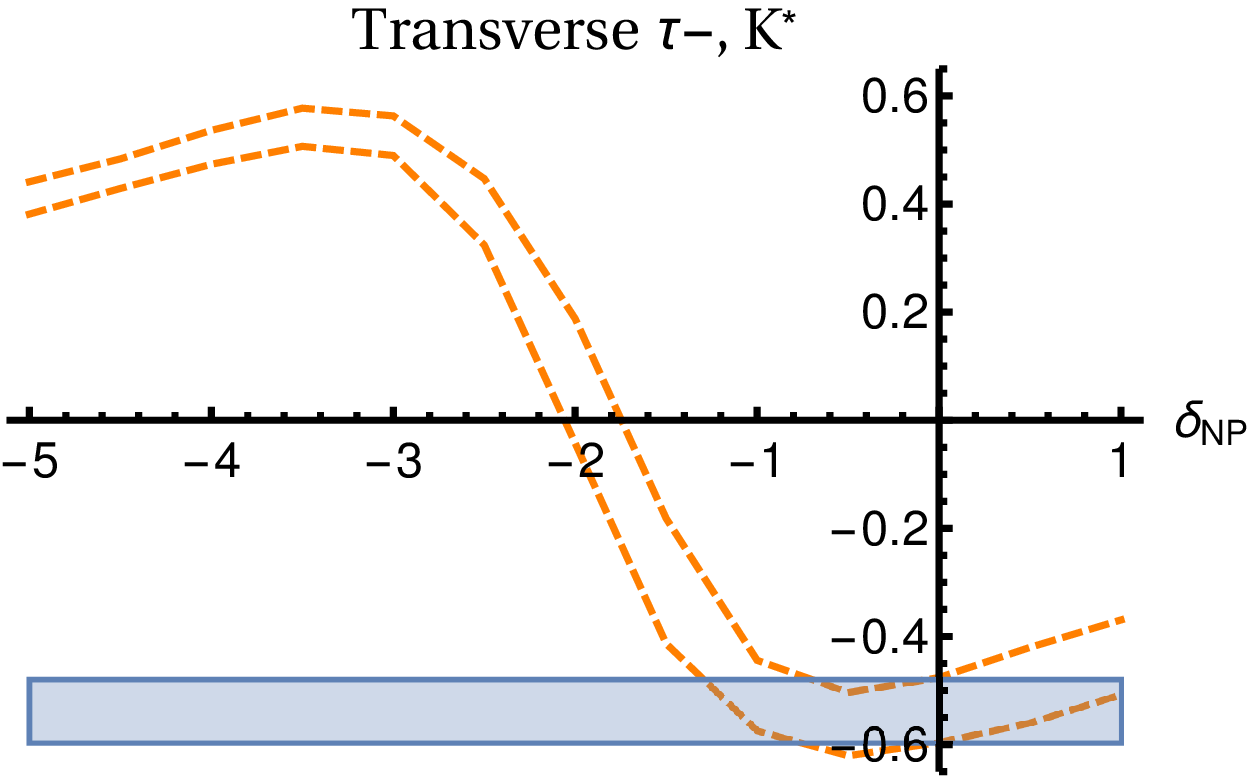}
	\caption{\it New physics effects in the form of a real $ \delta C_9 = - C'_9 = - \delta C_{10} = - C'_{10} \equiv \delta_{\rm NP} $, whose values are given in the horizontal axes. See Figure~\ref{fig:BtoKstarNPcase19} for more comments on the reading of the graphics. See Figure~\ref{fig:BtoKNPcase13} for plots concerning $ B \rightarrow K \tau^+ \tau^- $.}\label{fig:BtoKstarNPcase13}
\end{figure}

\begin{figure}[t]
\centering
	\includegraphics[scale=0.55]{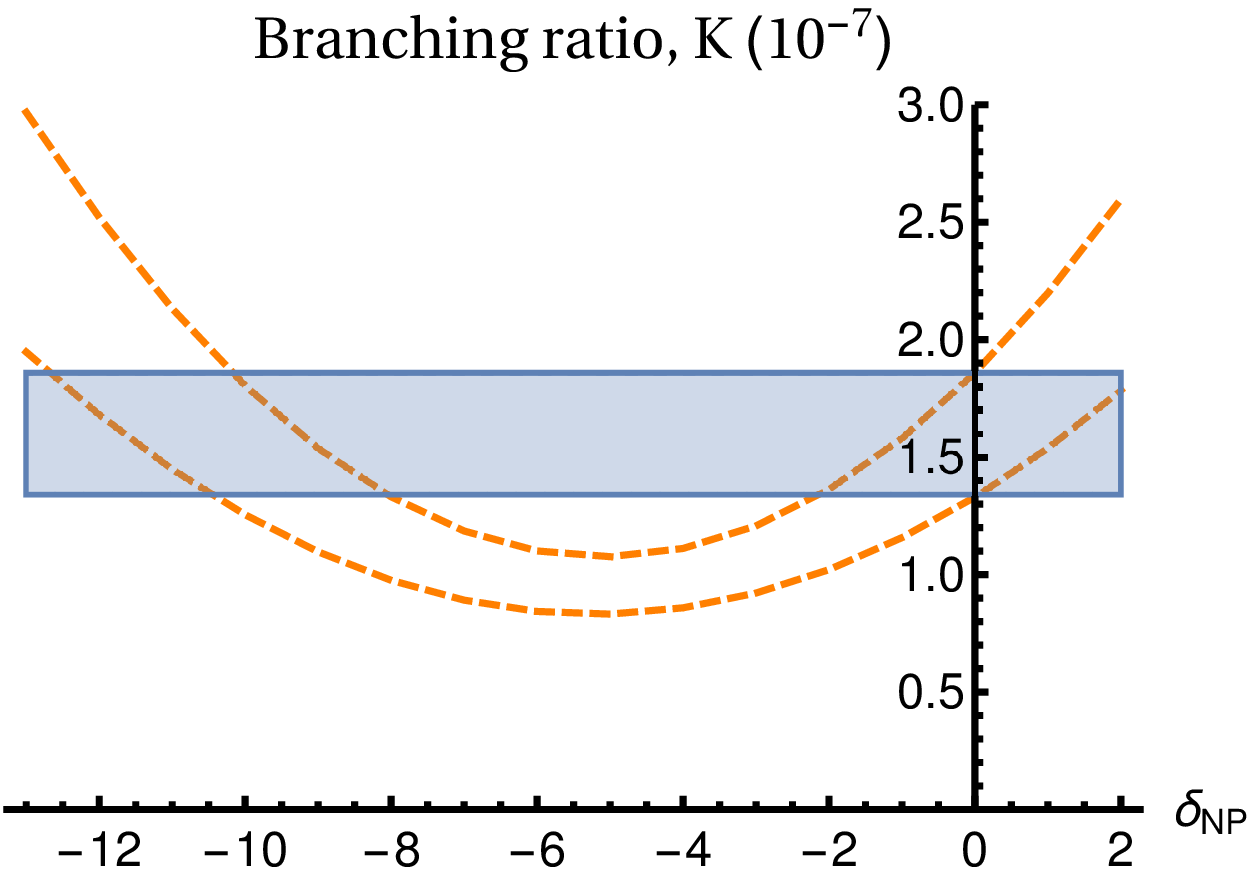}
	\hspace{3mm}\includegraphics[scale=0.55]{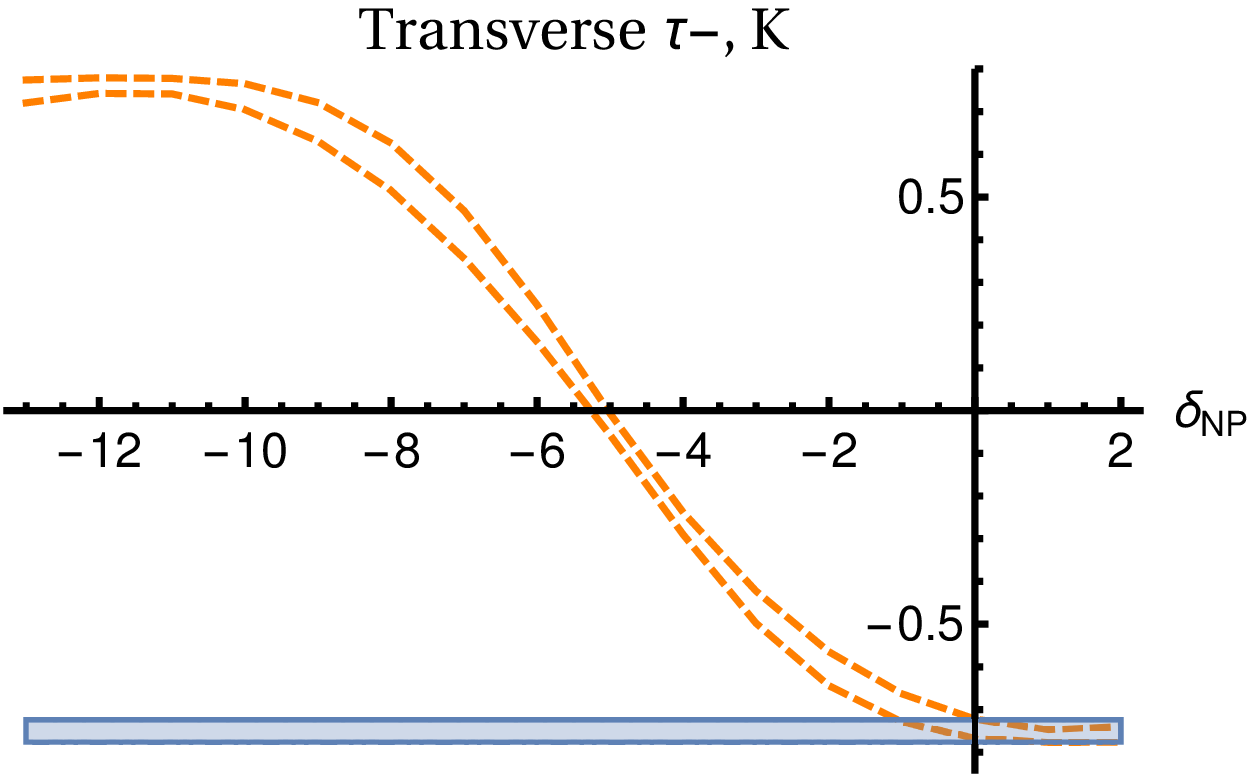} \\
	\vspace{5mm}
	\includegraphics[scale=0.55]{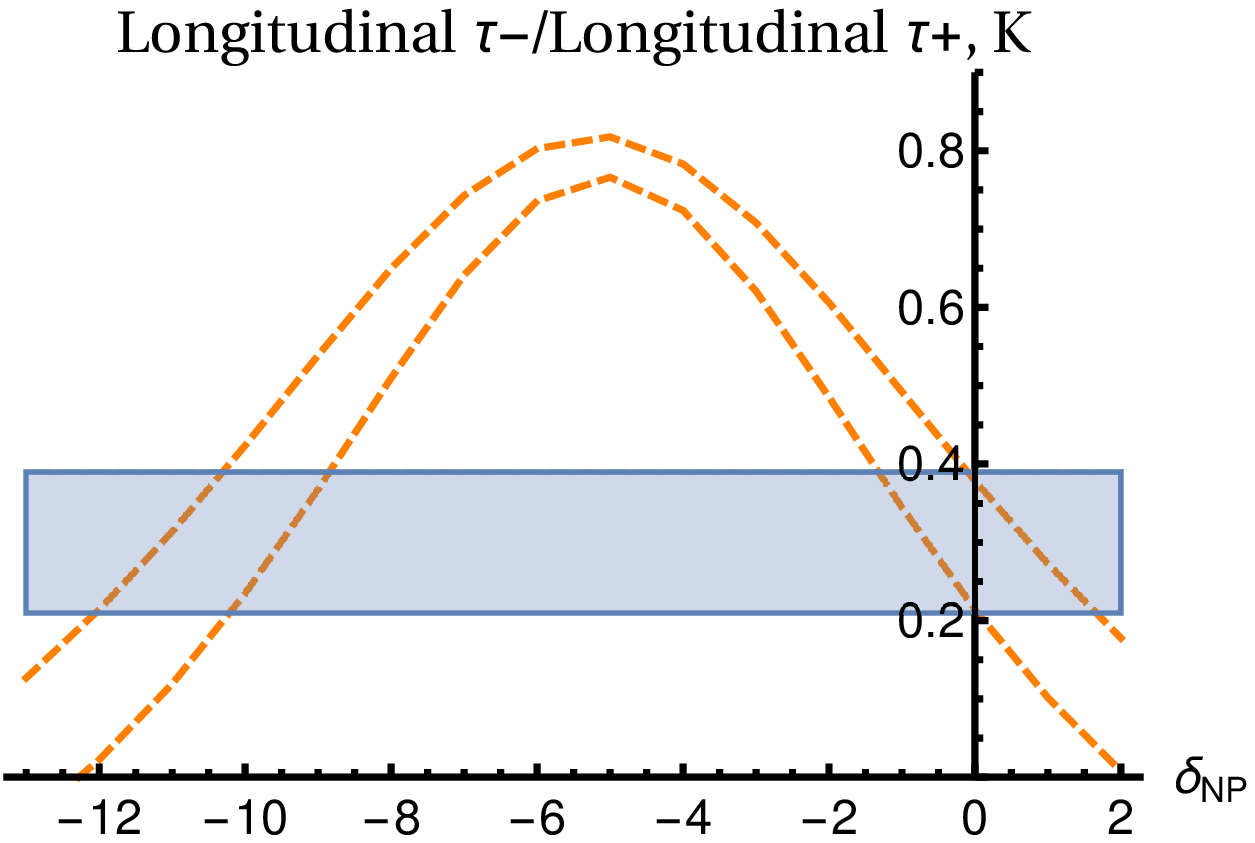}
	\hspace{3mm}\includegraphics[scale=0.55]{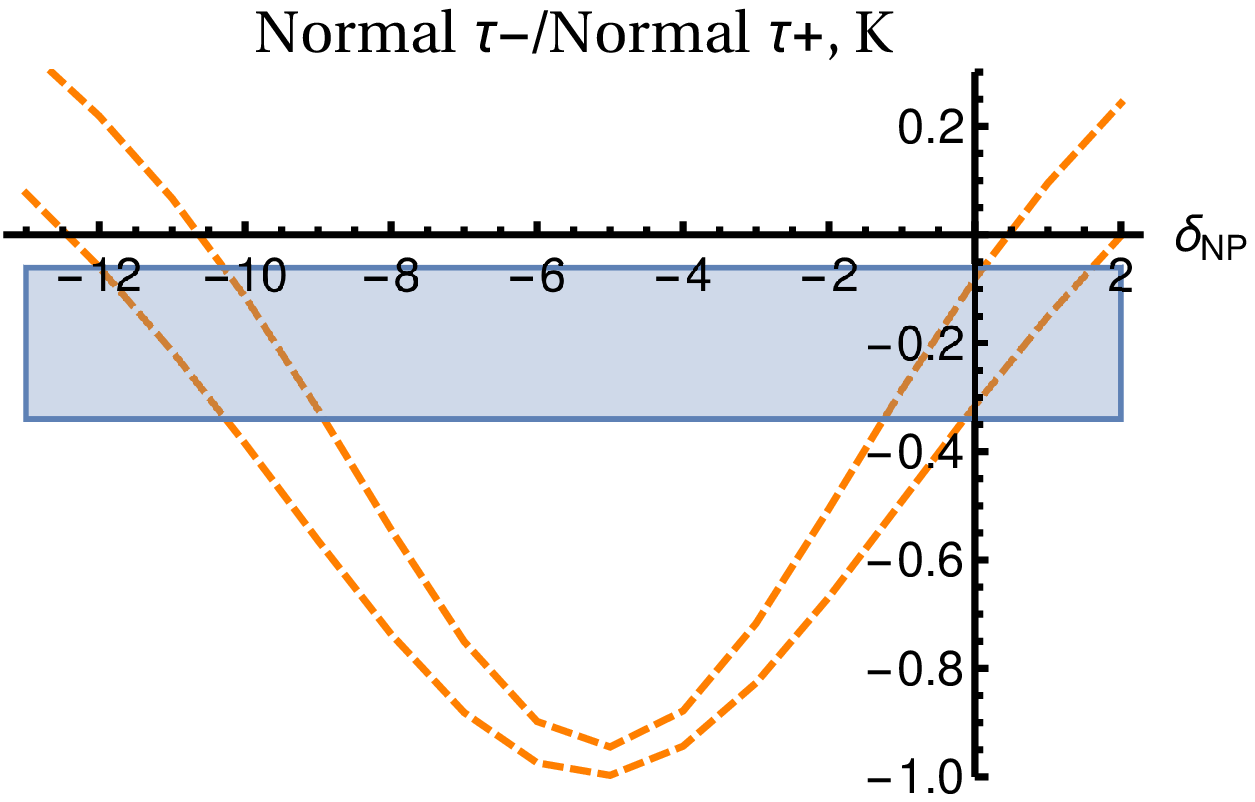}
	\caption{\it New physics effects in the form of a real $ \delta C_9 \equiv \delta_{\rm NP} $, whose values are given in the horizontal axes. The vertical axes give the values of the branching ratio, the transverse polarization of the $ \tau^- $, and the correlated longitudinal-longitudinal and normal-normal polarizations, in the SM (solid, filled blue, independent on the value of $ \delta_{\rm NP} $) and in the NP under consideration (dashed orange). The blue band corresponds to the errors seen in Table~\ref{tab:BtoKSMvalues}, which are recalculated for NP.}\label{fig:BtoKNPcase19}
\end{figure}

\begin{figure}[t]
\centering
	\includegraphics[scale=0.55]{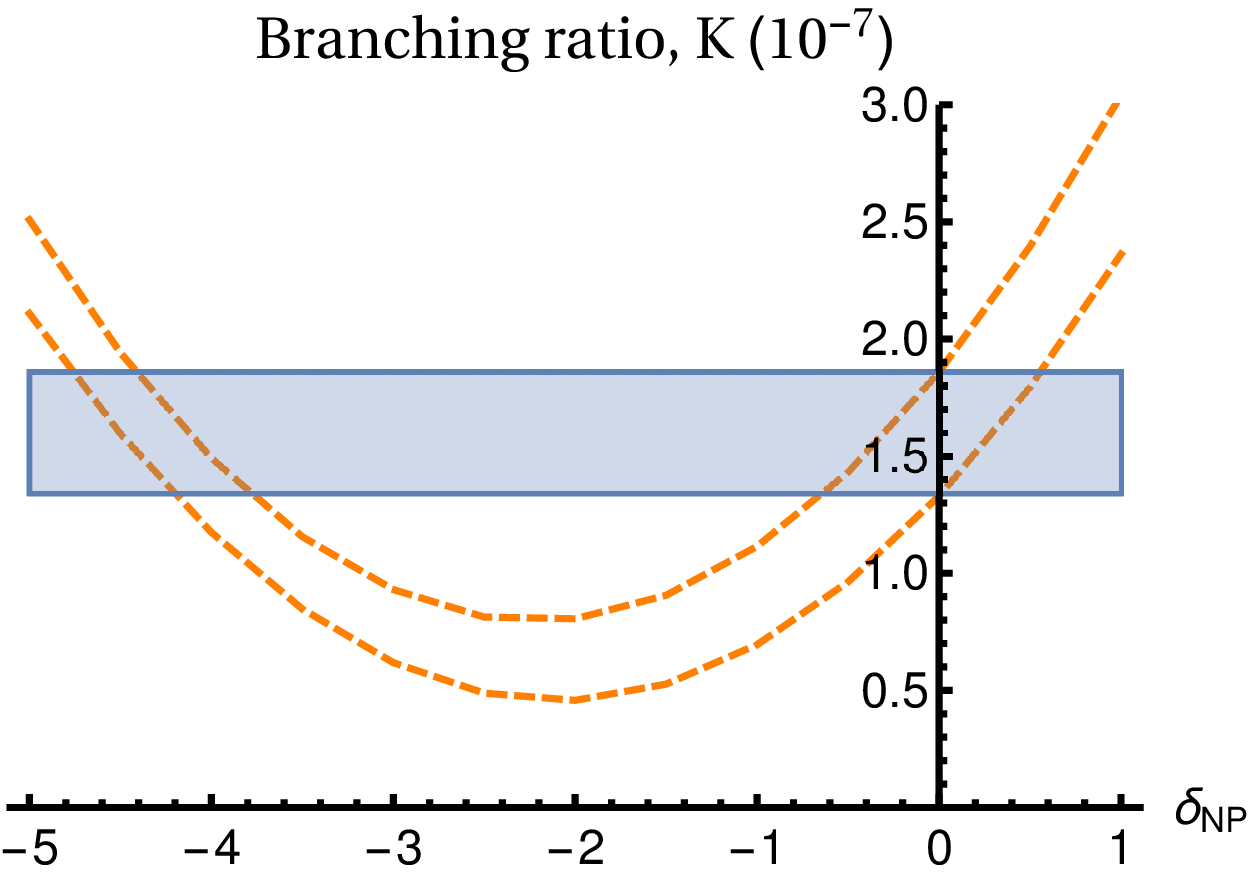}
	\hspace{3mm}\includegraphics[scale=0.55]{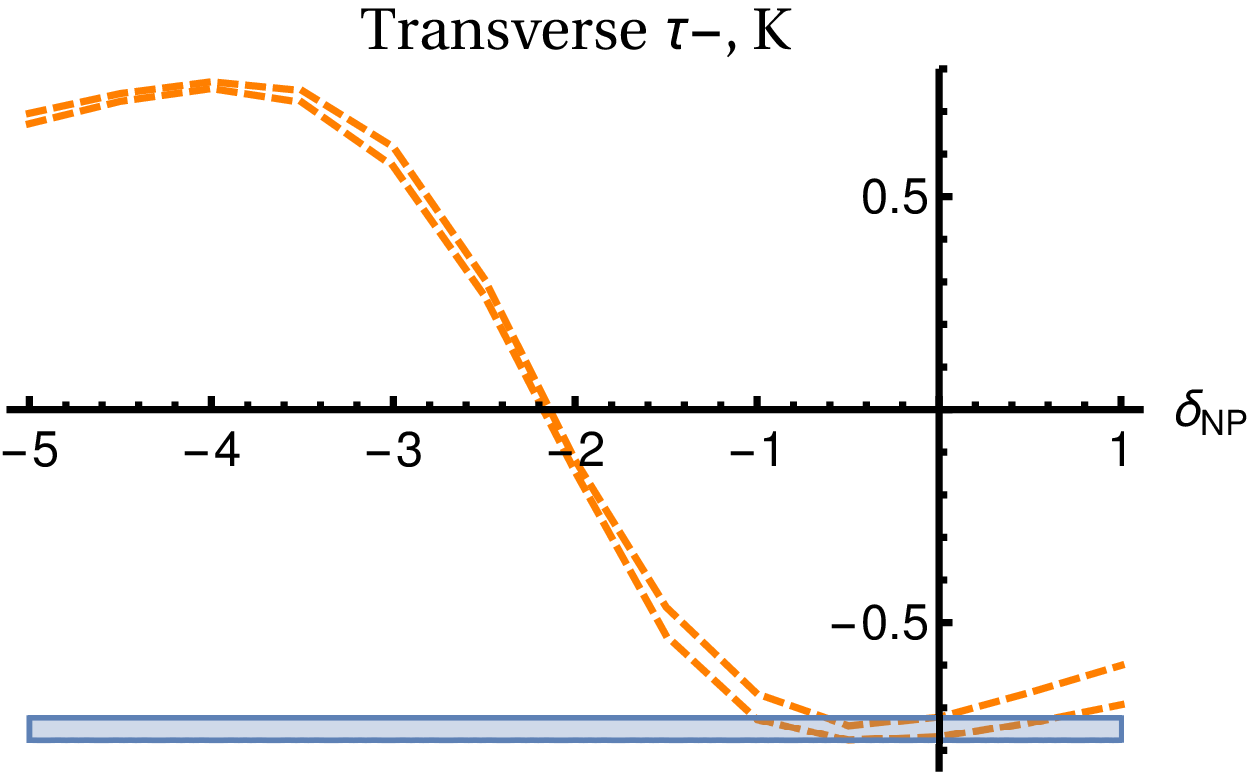} \\
	\vspace{5mm}
	\includegraphics[scale=0.55]{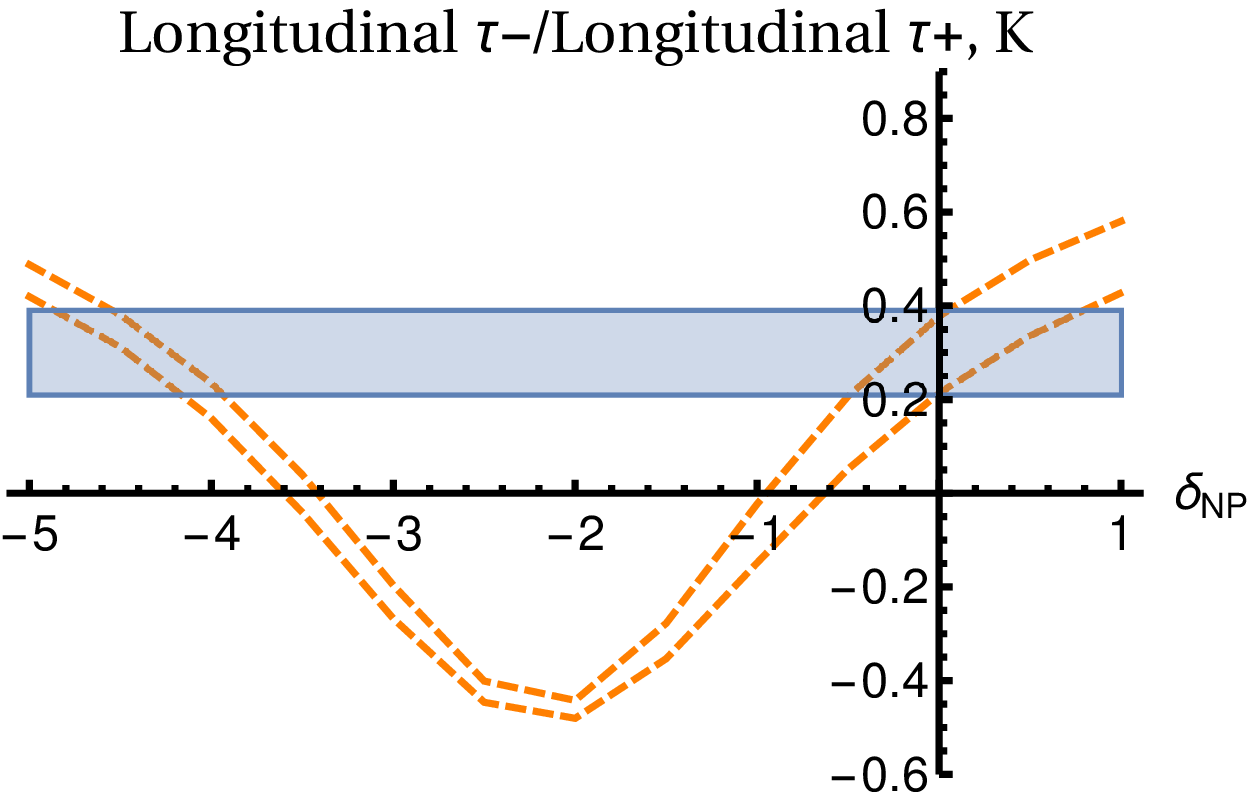}
	\hspace{3mm}\includegraphics[scale=0.55]{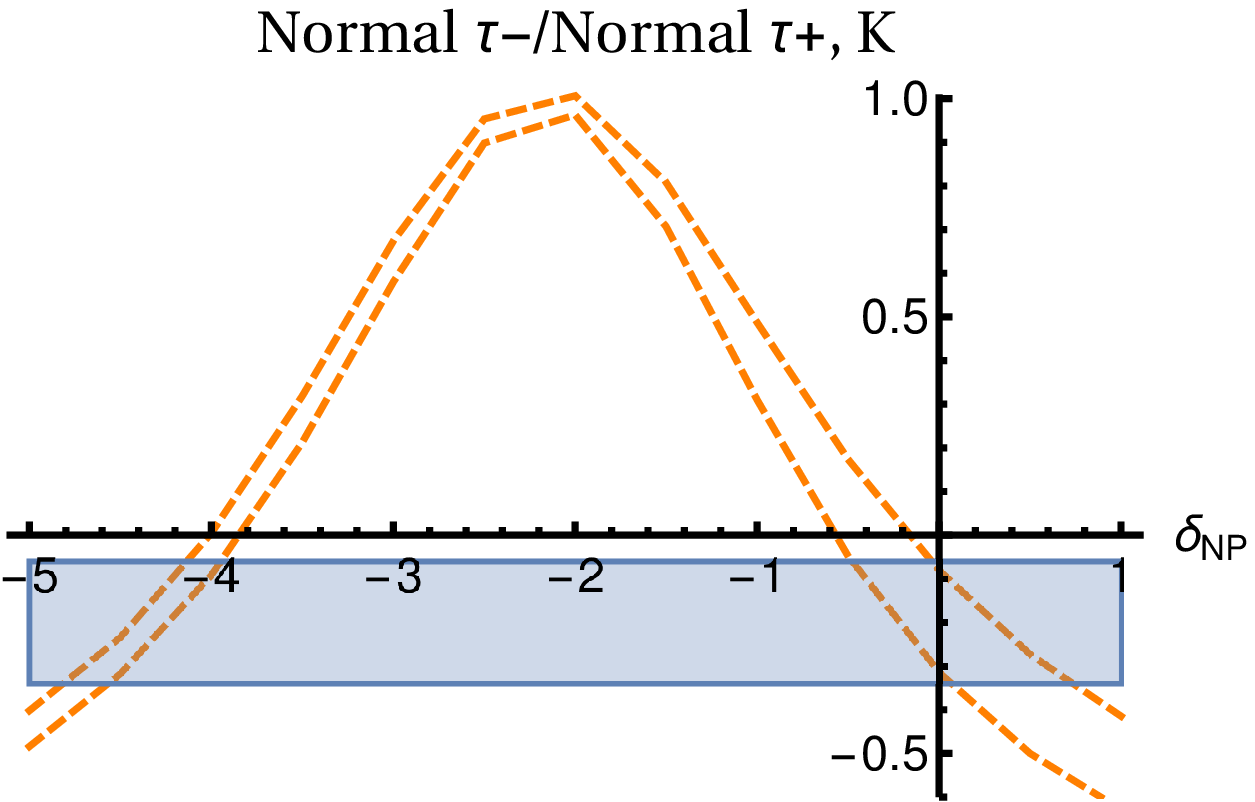}
	\caption{\it New physics effects in the form of a real $ \delta C_9 = - C'_9 = - \delta C_{10} = - C'_{10} \equiv \delta_{\rm NP} $, whose values are given in the horizontal axes. See Figure~\ref{fig:BtoKNPcase19} for more comments on the reading of the graphics.}\label{fig:BtoKNPcase13}
\end{figure}

In general, NP affecting $\tau$ polarization asymmetries can also be probed using more traditional observables such as the total rate or the various angular asymmetries.  A global sensitivity comparison is beyond the scope of our work. For the sake of illustration, we compare the sensitivities of the simplest and most accessible observables, namely the total rate and the forward-backward (FB) asymmetry. The FB asymmetry is defined as the branching ratio for $ \theta_\tau \in [0, \frac{\pi}{2}] $ minus the one for $ \theta_\tau \in [\frac{\pi}{2}, \pi] $ (see Figure~\ref{fig:ReferenceFrame}),
\begin{equation}\label{eq:FBasym}
\mathcal{A}_{FB} (q^2) = \frac{d \Gamma (z_\tau>0) / d q^2 - d \Gamma (z_\tau<0) / d q^2}{d \Gamma (z_\tau>0) / d q^2 + d \Gamma (z_\tau<0) / d q^2} \, .
\end{equation}
A broader class of FB asymmetries for the inclusive process $ B \rightarrow X_s \tau^+ \tau^- $ have been previously discussed in Ref.~\cite{Bensalem:2002ni}.

The operators $ O_9, O_{10}, O'_9, O'_{10}, O_7, O'_7 $ contribute at the tree level, and therefore deserve special attention when studying the impact of NP effects in $B_{(s)} \to M \ell^+\ell^-$ decays. Contributions from $ O_7 $ or the chirality flipped $ O'_7 $ are LFU and their non-standard effects are already severely constrained by existing measurements of rare $ B $ radiative and semi-electronic (semi-muonic) decay modes. We thus focus on the following illustrative NP scenarios 
\begin{eqnarray}\label{eq:cases}
{\rm \mathbf NP1 \, :} && \delta C_9 \; {\rm real} \, , \\
{\rm \mathbf NP2 \, :} && \delta C_9 = - C'_9 \; {\rm real} \, , \\
{\rm \mathbf NP3 \, :} && \delta C_9 = - C'_9 = - \delta C_{10} = - C'_{10} \; {\rm real} \, .
\end{eqnarray}
Scenarios \textbf{NP1} and \textbf{NP2} for $ \delta C^{\mu \mu}_9 \approx -1 $ and $ \delta C^{\mu \mu}_9 = - C^{' \mu \mu}_9 \approx -0.9 $, respectively, are actually favored by current anomalies in muon data \cite{Descotes-Genon:2015uva, Altmannshofer:2013foa}. Scenario \textbf{NP3} for $ C^{\mu \mu}_9 = - C^{' \mu \mu}_9 = - \delta C^{\mu \mu}_{10} = - C^{' \mu \mu}_{10} \approx -0.7 $ is also favored in some analyses \cite{Descotes-Genon:2015uva}. Later in the text, we will also comment on other motivated scenarios such as with real $\delta C_9 = - \delta C_{10}$ or with complex $C_{9,10}^{(\prime)}$.
In the following, we focus on the $ B \rightarrow K^\ast \tau^+ \tau^- $ mode, which was also our subject in Section~\ref{sec:exp}.

Since we are interested in cases where the polarization asymmetries exhibit better NP sensitivity than the branching ratio, we choose the ranges of variation for the NP contributions such that the branching ratio changes by at most a factor $ \approx 2 $. Note that existing direct bounds on $ B^+ \rightarrow K^+ \tau^+ \tau^- $~\cite{Bobeth:2011st} and $ B_s \rightarrow \tau^+ \tau^- $~\cite{Aaij:2017xqt} constrain NP effects to $ | C^{\rm NP}_{mn} | \lesssim 10^3 $, where $ C^{\rm NP}_{mn} $ is the NP Wilson coefficient of the operator $ (\bar{s} \gamma^\mu P_m b) \, (\bar{\tau} \gamma_\mu P_n \tau) $, $ m, n = L, R $ (see also \cite{Grossman:1996qj,Dighe:2010nj} for indirect bounds). In addition, for heavy NP respecting SM $ SU(2)_L $ gauge invariance, existing bounds on $ B \rightarrow K^{(\ast)} \nu \bar{\nu} $ constrain $ | C^{\rm NP}_{mL} | \lesssim \mathcal{O} (10) $, $ m = L, R $~\cite{Alonso:2015sja}. In the NP ranges considered here, none of these existing bounds is violated. 

In Figures~\ref{fig:BtoKstarNPcase19}-\ref{fig:BtoKNPcase13} we present the NP induced variations in the branching ratio together with the asymmetries for which we have the most notable effects in NP scenarios \textbf{NP1}, \textbf{NP2}, \textbf{NP3} (for \textbf{NP2} the values for the process $ B \rightarrow K $ are unchanged with respect to the SM, and therefore not shown). Similar plots are obtained for the decay $ B_s \rightarrow \phi $ and are not displayed here. In the figures, the horizontal blue bands represent the SM predictions, with ``Charm", FF and WC uncertainty contributions combined linearly. In  presence of the NP manifestations considered here, $ | \mathcal{P}_L^{-} | = | \mathcal{P}_L^{+} | $, $ | \mathcal{P}_T^- (K) | = | \mathcal{P}_T^+ (K) | $, $ | \mathcal{P}_N^{-} | = | \mathcal{P}_N^{+} | $, $ | \mathcal{P}_{LT} (K) | = | \mathcal{P}_{TL} (K) | $, $ | \mathcal{P}_{LN} | = | \mathcal{P}_{NL} | $, $ | \mathcal{P}_{TN} | = | \mathcal{P}_{NT} | $ and $ \mathcal{A}_{FB} (K) = 0 $.

We highlight $ \langle \mathcal{P}_L^{\pm} (V) \rangle $ whose SM-like values would exclude large regions of the NP parameter space. Moreover, we note that some SM-like branching ratio solutions actually flip the sign of  $ \langle \mathcal{P}_L^{\pm} (V) \rangle $. To illustrate the use of the longitudinal asymmetry to discriminate models, note from Figures~\ref{fig:BtoKstarNPcase19} and \ref{fig:BtoKstarNPcase9} that though the longitudinal asymmetries for \textbf{NP1} and \textbf{NP2} have similar shapes, they evolve differently with the value of $ \delta_{\rm NP} $, which parameterizes the size of NP effects: for instance, around $ \delta_{\rm NP} \approx -2 $ the value of $ \langle \mathcal{P}^{-}_L (V) \rangle $ for the case \textbf{NP2} is very different compared to the SM, while that is not true for \textbf{NP1}.

Apart from the longitudinal polarization, the FB asymmetry and the transverse polarization may also show sign flips, for both $ B \rightarrow K^\ast \tau^+ \tau^- $ and $ B \rightarrow K \tau^+ \tau^- $. Though the FB asymmetry is not enhanced, it is useful for distinguishing cases \textbf{NP2} and \textbf{NP3}, since as seen from Figures~\ref{fig:BtoKstarNPcase9} and \ref{fig:BtoKstarNPcase13} the branching ratios and $ \langle \mathcal{P}_L^{\pm} (V) \rangle $, $ \langle \mathcal{P}_T^{-} (V) \rangle $ have similar shapes. Moreover, \textbf{NP2} and \textbf{NP3} can also be distinguished by looking at $ B \rightarrow K \tau^+ \tau^- $, cf. Figure~\ref{fig:BtoKNPcase13}, since for \textbf{NP2} all the observables have SM-like values for this channel.

\begin{figure}
\centering
	\includegraphics[scale=0.55]{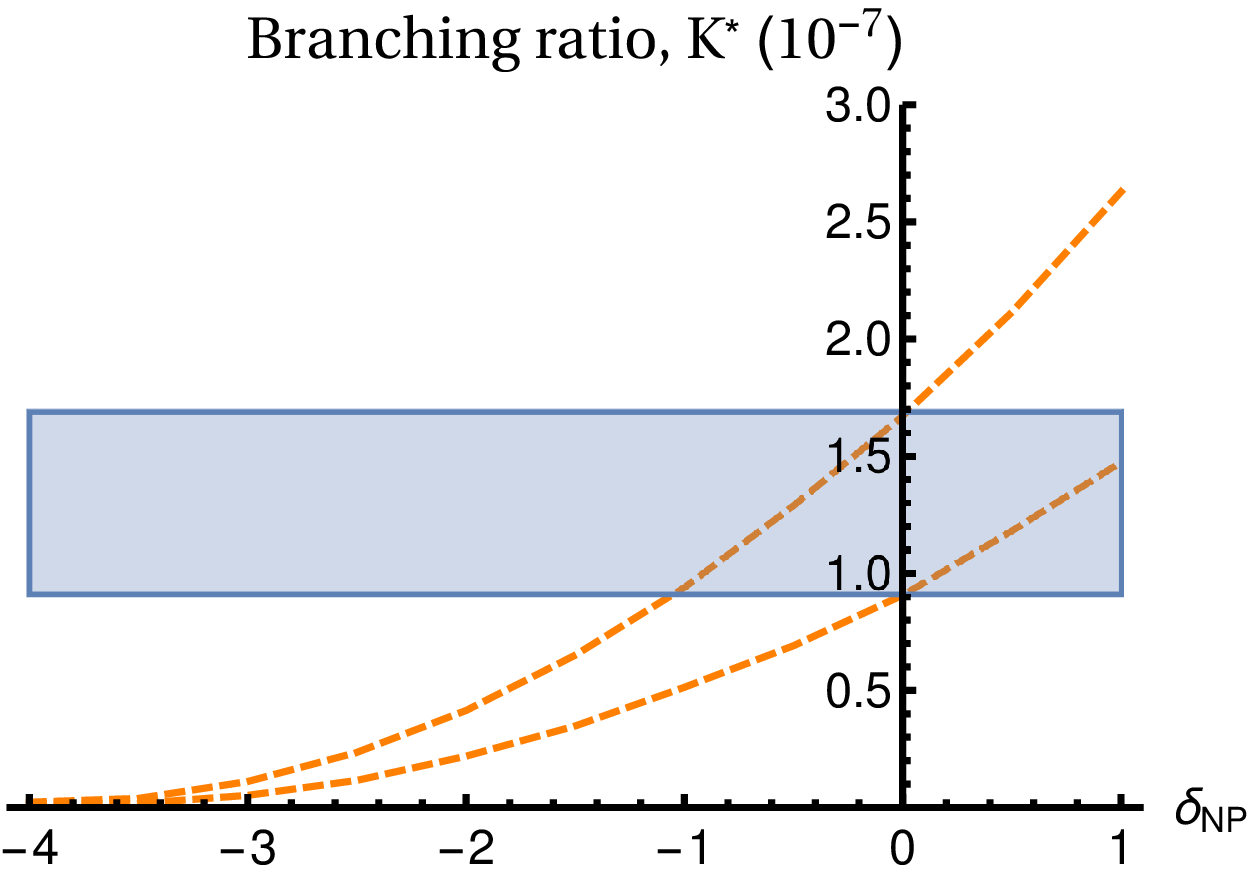}
	\hspace{3mm}\includegraphics[scale=0.55]{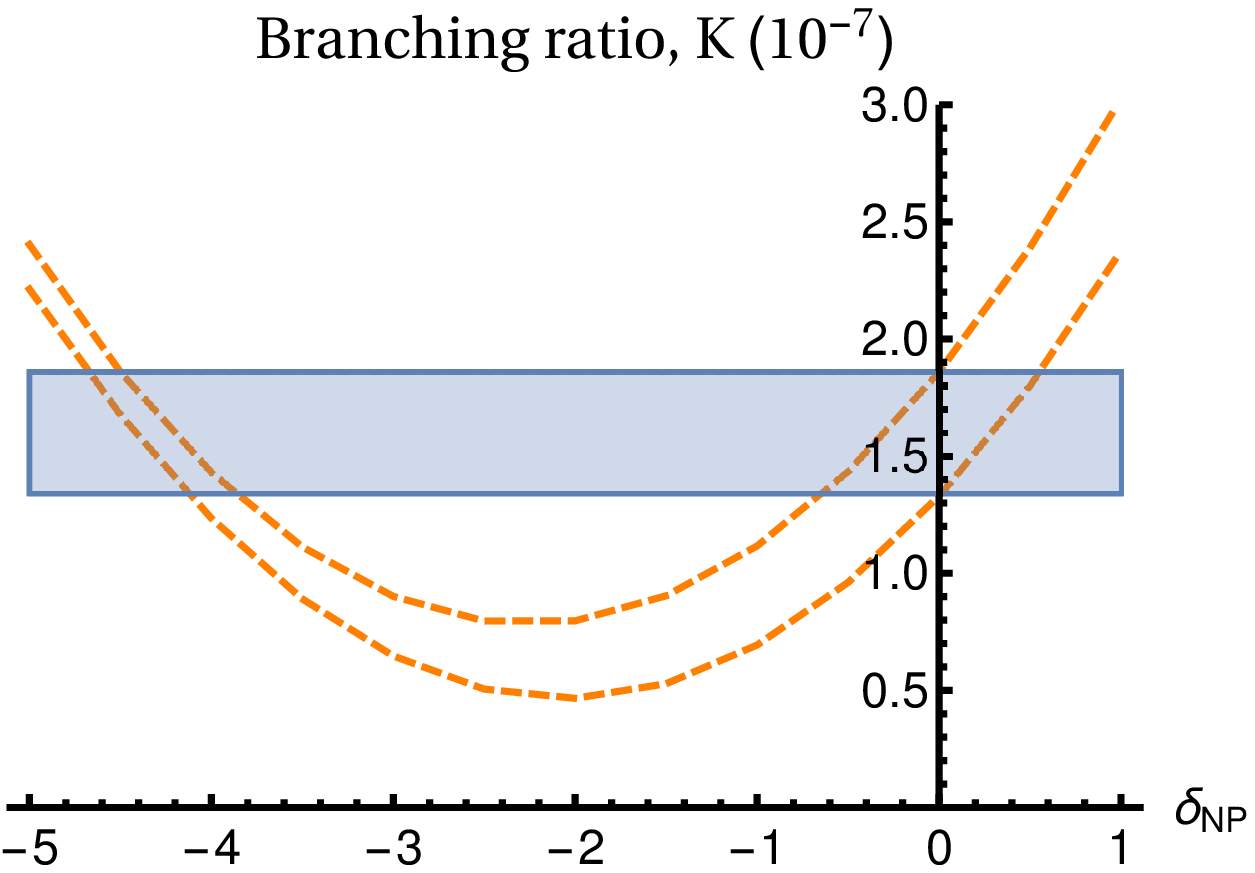}
	\caption{\it New physics effects in the form of a real $ \delta C_9 = - \delta C_{10} \equiv \delta_{\rm NP} $, whose values are given in the horizontal axes. See Figs.~\ref{fig:BtoKstarNPcase19}, \ref{fig:BtoKNPcase19} for more comments on the reading of the graphics.}\label{fig:BtoKandBtoKstarNPcase12}
\end{figure}

Considering other NP scenarios, note from Figure~\ref{fig:BtoKandBtoKstarNPcase12} that the case of a real $ \delta C_9 = - \delta C_{10} $ (also favored by muon data for $ C^{\mu \mu}_9 = - \delta C^{\mu \mu}_{10} \approx -0.7 $ \cite{Descotes-Genon:2015uva}) shows a large suppression of the branching ratio $ \langle BR (K^\ast) \rangle $. For obvious reasons then, if this case is realized for a large interval of the NP shifts to the Wilson coefficients, $ B \rightarrow K^\ast \tau^+ \tau^- $ asymmetries will probably be out of reach experimentally.

\begin{figure}
\centering
	\includegraphics[scale=0.55]{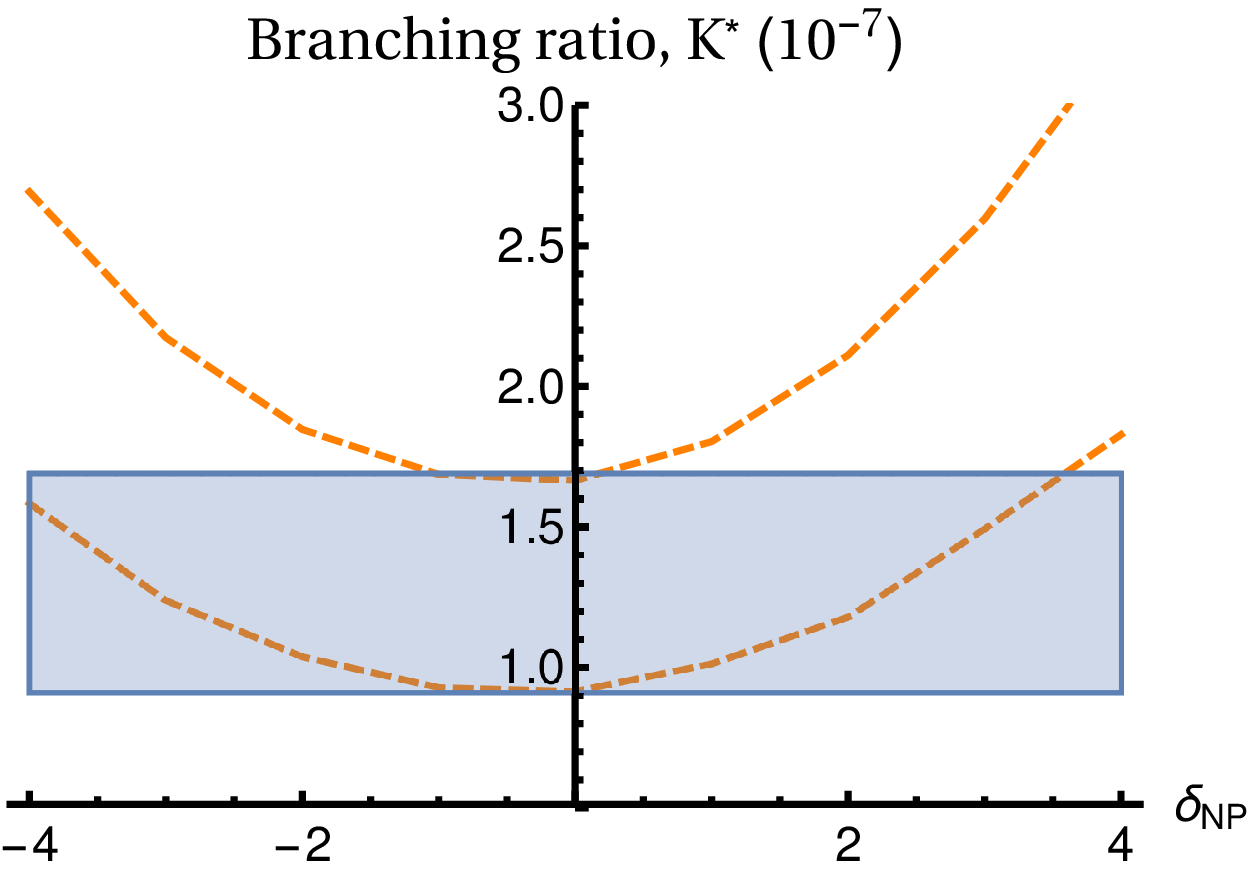}
	\hspace{3mm}\includegraphics[scale=0.55]{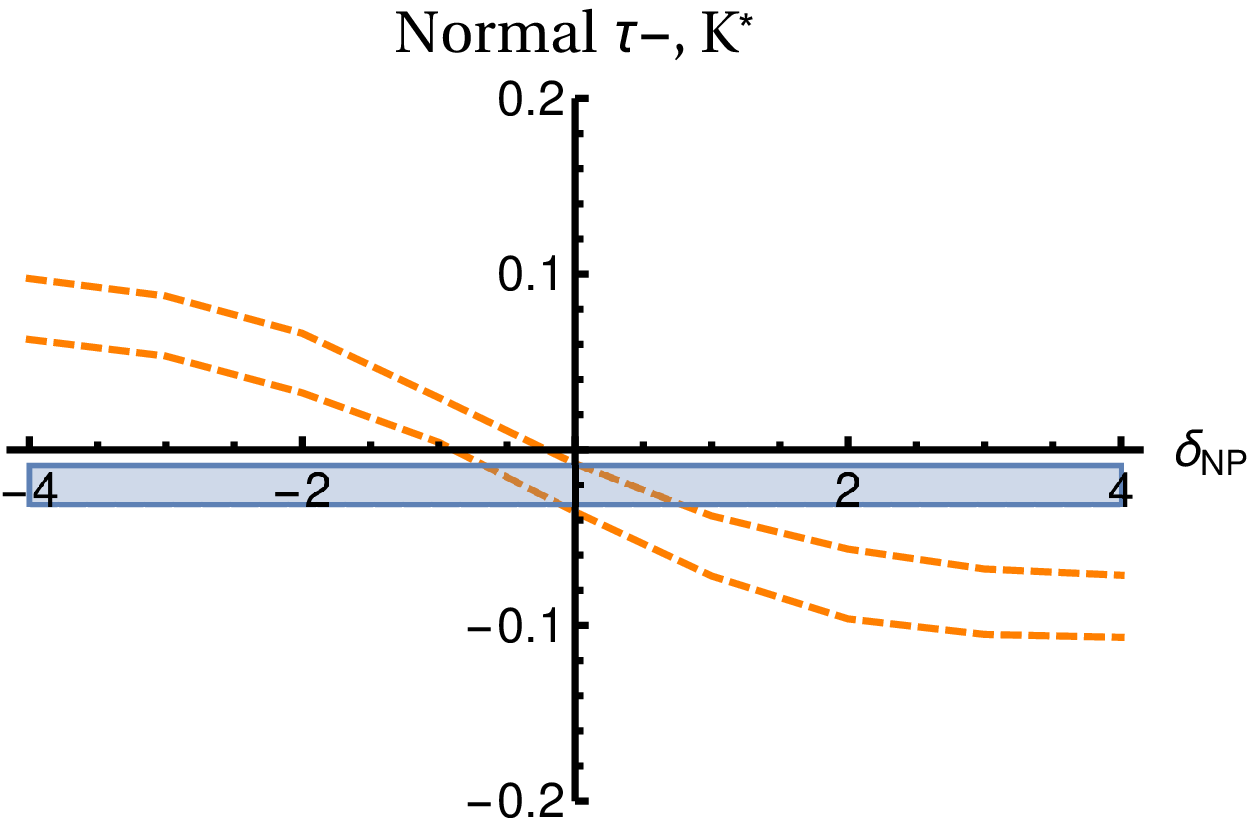}
	\caption{\it New physics effects in the form of an imaginary $ \delta C_9 \equiv i \, \delta_{\rm NP} $. The values for the real $ \delta_{\rm NP} $ are given in the horizontal axes. See Figure~\ref{fig:BtoKstarNPcase19} for more comments on the reading of the graphics.}\label{fig:imaginaryc9}
\end{figure}

\begin{figure}
\centering
	\includegraphics[scale=0.55]{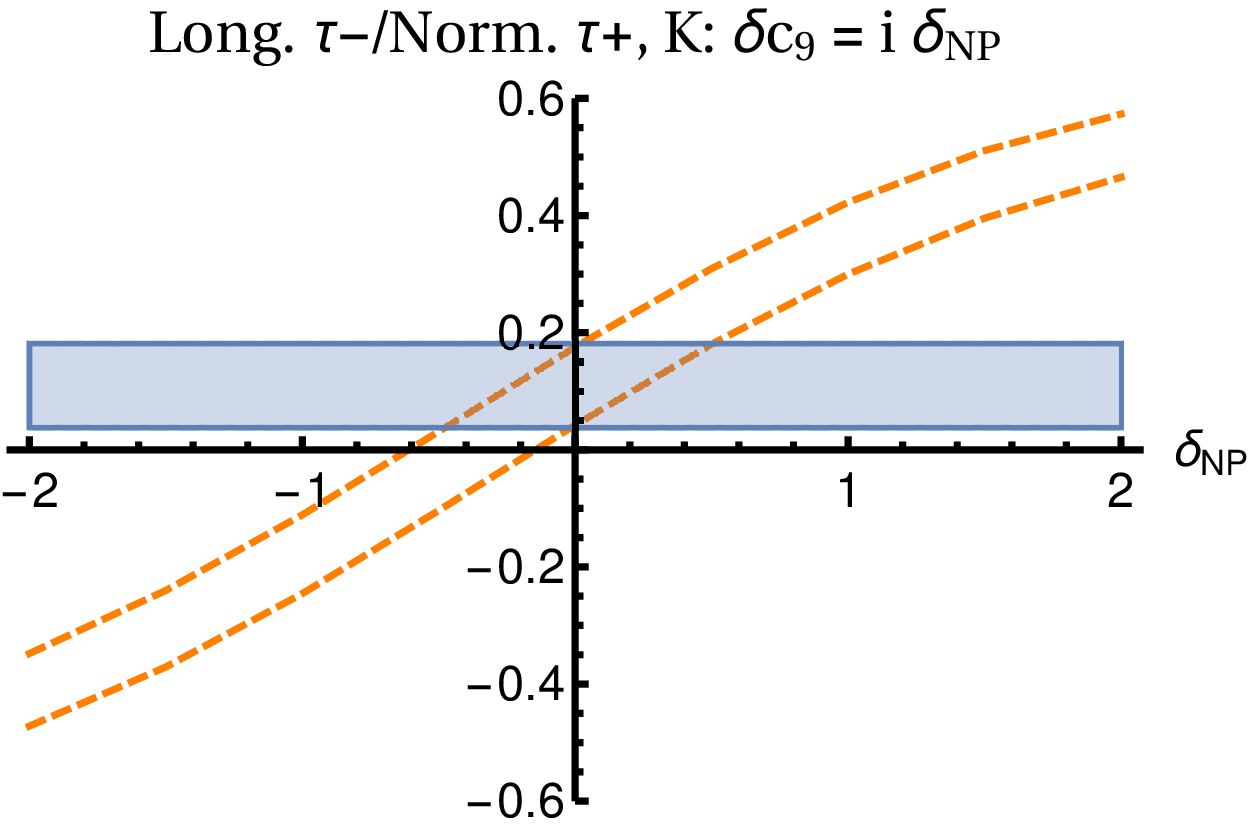}
	\hspace{3mm}\includegraphics[scale=0.55]{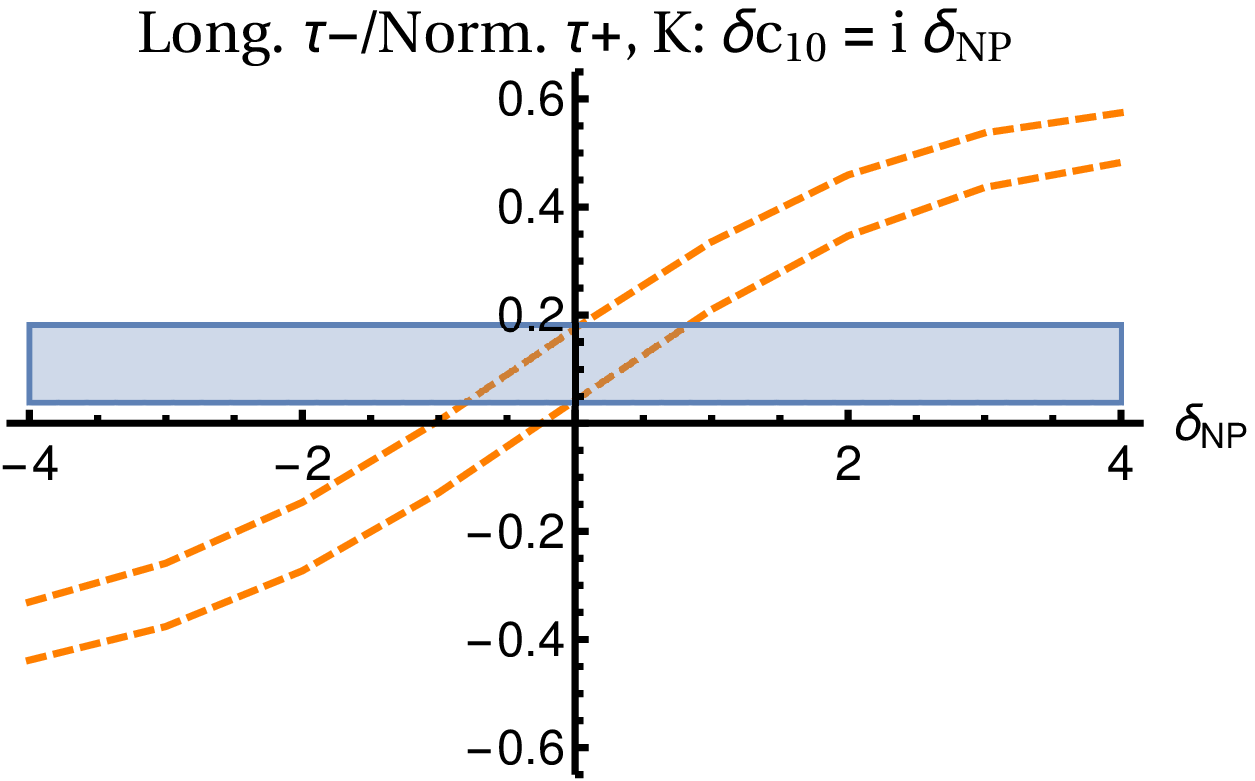}
	\caption{\it New physics effects coming as $ \delta C_9 \equiv i \, \delta_{\rm NP} $ (left), or $ \delta C_{10} \equiv i \, \delta_{\rm NP} $ (right). See Figure~\ref{fig:BtoKNPcase19} for more comments on the reading of the graphics.}\label{fig:LNayms}
\end{figure}

Imaginary phases have also been considered in the context of muon data anomalies in, for instance, Refs.~\cite{Altmannshofer:2012az, Altmannshofer:2013foa} and the analyses indicate possible large imaginary components. 
Such an example is illustrated in Figure~\ref{fig:imaginaryc9} for imaginary $ \delta C_9 $. We observe that no large effects in single tau polarization asymmetries are induced in this case (similar comments for $ \langle \mathcal{P}^{-}_N (V) \rangle $ would also apply for an imaginary $ \delta C_{10} $, and moreover imaginary values for $ C'_9, C'_{10} $).\footnote{Note however that the tensorial operator $ ( \bar{s} \sigma_{\alpha \beta} b ) \, \epsilon^{\alpha \beta \mu \nu} \, ( \bar{\tau} \sigma_{\mu \nu} \tau ) $ not considered here can significantly enhance the normal polarization.} A significant enhancement is only observed in $ \langle \mathcal{P}_{LN} (K) \rangle $, see Figure~\ref{fig:LNayms}. In any case, observables other than the tau polarizations may be more sensitive to complex phases, see e.g. Ref.~\cite{Becirevic:2011bp}.

To conclude, in Table~\ref{tab:enhancedAsyms} we gather the asymmetries that have the largest values in presence of the NP scenarios considered here. For instance, scenarios \textbf{NP1} and \textbf{NP2} can be distinguished by an enhanced value of $ \langle \mathcal{P}_T^{+} (V) \rangle $, while different values of $ \delta C_9 $ in scenario \textbf{NP1} can be distinguished by measuring an anomalously enhanced value of $ \langle \mathcal{P}_T^{+} (V) \rangle $, $ \langle \mathcal{P}_{LT} (V) \rangle $, $ \langle \mathcal{P}_{TT} (V) \rangle $, $ \langle \mathcal{P}_{TL} (V) \rangle $, or a SM-like $ \langle \mathcal{P}_L^{\pm} (V) \rangle $. Of course, a combined correlated measurement of the various tau polarization asymmetries considered here can in principle provide even more powerful tests and discriminants between the SM and various NP scenarios. 

\begin{table}[t]
	\centering
	\begin{tabular}{cc}
	\hline
	$ \langle \mathcal{P}_L^{-} (V) \rangle \approx -0.5 $, & \\
	$ \langle \mathcal{P}_L^{+} (V) \rangle \approx {+}0.5 $, & SM-like \\
	$ \langle \mathcal{P}_T^{-} (V) \rangle \approx -0.5 $ & \\
	\hline
	\hline
	 & $ \delta C_9 \lesssim -5 $, \\
	$ \langle \mathcal{P}_L^{-} (V) \rangle \approx +0.5 $, & $ \delta C_9 = - C'_9 \lesssim -3 $, \\
	$ \langle \mathcal{P}_L^{+} (V) \rangle \approx {-}0.5 $ & $ \delta C_9 = - C'_9 = $ \\
	 & $ - \delta C_{10} = - C'_{10} \lesssim -3 $ \\
	\hline
	 & $ \delta C_9 = - C'_9 \lesssim -3 $, \\
	$ \langle \mathcal{P}_T^{-} (V) \rangle \approx +0.5 $ & $ \delta C_9 = - C'_9 = $ \\
	 & $ - \delta C_{10} = - C'_{10} \lesssim -3 $ \\
	\hline
	$ \langle \mathcal{P}_T^{+} (V) \rangle \approx {{-}}0.5 $ & $ \delta C_9 \lesssim -6 $ \\
	\hline
	 & $ \delta C_9 \approx -3 $, \\
	$ \langle \mathcal{P}_{TL} (V) \rangle \approx {-} 0.5 $ & $ \delta C_9 = - C'_9 \approx -2 $, \\
	 & $ \delta C_9 = - C'_9 = $ \\
	 & $ - \delta C_{10} = - C'_{10} \approx -2 $ \\
	\hline
	$ \langle \mathcal{P}_{LT} (V) \rangle \approx {{-}} 0.5 $ & $ \delta C_9 \approx -5 $ \\
	\hline
	$ \langle \mathcal{P}_{TT} (V) \rangle \approx {{-}} 0.4 $ & $ \delta C_9 \approx -4 $ \\
	\hline
	 & $ \delta C_9 \approx -4 $, \\
	$ \langle \mathcal{P}_{NN} (V) \rangle \approx {}{{-}} 0.4 $ & $ \delta C_9 = - C'_9 \approx -2 $, \\
	 & $ \delta C_9 = - C'_9 = $ \\
	 & $ - \delta C_{10} = - C'_{10} \approx -2 $ \\
	\hline
	\end{tabular}
	\quad
	\begin{tabular}{cc}
	\hline
	$ \langle \mathcal{P}_T^- (K) \rangle \approx -0.8 $, & \\
	$ \langle \mathcal{P}_T^+ (K) \rangle \approx {{+}}0.8 $, & SM-like \\
	$ \langle \mathcal{P}_{TT} (K) \rangle \approx {{-}}0.7 $ & \\
	\hline
	\hline
	$ \langle \mathcal{P}_T^- (K) \rangle \approx +0.8 $ & $ \delta C_9 \lesssim -8 $, \\
	$ \langle \mathcal{P}_T^+ (K) \rangle \approx {{-}}0.8 $ & $ \delta C_9 = - C'_9 = $ \\
	 & $ - \delta C_{10} = - C'_{10} \lesssim -3 $ \\
	\hline
	$ \langle \mathcal{P}_{TL} (K) \rangle \approx {-} 0.5 $, & $ \delta C_9 \approx -5 $ \\
	$ \langle \mathcal{P}_{LT} (K) \rangle \approx {{-}} 0.5 $ & \\
	\hline
	$ \langle \mathcal{P}_{TT} (K) \rangle \approx {{-}} 0.8 $ & $ \delta C_9 \approx -5 $ \\
	\hline
	$ \langle \mathcal{P}_{TT} (K) \rangle \approx {{-}} 0.4 $ & $ \delta C_9 = - C'_9 = $ \\
	 & $ - \delta C_{10} = - C'_{10} \lesssim -2 $ \\
	\hline
	$ \langle \mathcal{P}_{NN} (K) \rangle \approx {}{{-}} 1 $ & $ \delta C_9 \approx -5 $ \\
	\hline
	$ \langle \mathcal{P}_{NN} (K) \rangle \approx {}{{+}} 1 $ & $ \delta C_9 = - C'_9 = $ \\
	 & $ - \delta C_{10} = - C'_{10} \lesssim -2 $ \\
	\hline
	$ \langle \mathcal{P}_{LL} (K) \rangle \approx {+} 0.8 $ & $ \delta C_9 \approx -5 $ \\
	\hline
	$ \langle \mathcal{P}_{LL} (K) \rangle \approx {-} 0.4 $ & $ \delta C_9 = - C'_9 = $ \\
	 & $ - \delta C_{10} = - C'_{10} \lesssim -2 $ \\
	\hline
	\end{tabular}
	\caption{\it Non-exhaustive table making the correspondence between the observation of an asymmetry at or beyond the level $ \pm 0.4 $, and the physical scenarios among \textbf{NP1}, \textbf{NP2}, \textbf{NP3} this observation implies. Above, $ V = K^\ast, \phi $. Of course, the non-observation of $ \langle \mathcal{P}_L^{\pm} (V) \rangle $, $ \langle \mathcal{P}_T^{-} (V) \rangle $, $ \langle \mathcal{P}_T^\pm (K) \rangle $, $ \langle \mathcal{P}_{TT} (K) \rangle $ at a sizable level also implies the existence of physics beyond the SM, which is not discussed in this table.}\label{tab:enhancedAsyms}
\end{table}

%
\section{Conclusions}
\label{sec:Conclusions}
%

In light of the recent intriguing experimental results indicating possible LFU violations in $B\to K^{(*)} \ell^+ \ell^-$ decays, it is important to investigate possible NP effects in rare semi-tauonic $b$-hadron decays. Currently these modes are only poorly constrained from the experimental side, allowing for potentially large NP effects. Given the difficulty to reconstruct the $ B_{(s)} \rightarrow M \tau^+ \tau^- $ decays, where $ M $ is a pseudoscalar or vector meson, we have argued that a future high energy $ e^+ e^- $ collider could play a crucial role in gaining experimental access to these rare processes.

In particular, working with the current FCC-$ee$ collider and detector design parameters we have presented a case for the viability of a full reconstruction of the $ B^0 \rightarrow K^{\ast 0} \tau^+ \tau^- $ decays in $e^+ e^-$ collisions at the $Z$-resonance mass. Our results indicate that of the order of a few thousands fully reconstructed events can be expected at the baseline collider luminosity. Based on these encouraging results we have considered observables other than the branching ratio. In particular, we have investigated the lepton polarization asymmetries, which are inaccessible in light lepton modes and thus provide uniquely new probes of NP in rare semileptonic $b$-hadron decays. Investigating closely the different sources of theoretical uncertainty we have provided precise SM predictions for these observables in  $ B \rightarrow K^{(\ast)} \tau^+ \tau^- $ decays and investigated their sensitivity to some motivated and representative NP scenarios. In particular, the single longitudinal and transverse asymmetries of the taus in the $ B \rightarrow K^{\ast} \tau^+ \tau^- $ decays are particularly sensitive to the NP scenarios considered here, while for the $ B \rightarrow K \tau^+ \tau^- $ decay the correlated longitudinal-longitudinal and normal-normal asymmetries may show sizable enhancements. Table~\ref{tab:enhancedAsyms} summarizes our findings on the possible enhancement or modulation of a variety of tau polarization asymmetries. For completeness we have computed these observables in the SM also for the $ B_s / \bar{B}_s \rightarrow \phi \tau^+ \tau^- $ decays which are not self-tagging and thus seem less suitable for experimental tau polarization studies. 

Further dedicated experimental studies will certainly be required in order to firmly establish the NP sensitivity of tau polarization observables in rare semi-tauonic $ B $ decays in the context of the FCC-$ee$. More generally however, we hope our preliminary results will help to strengthen the (flavor) physics case for the FCC-$ee$.

\section*{Acknowledgments}
{JFK and LVS thank Wolfgang Altmannshofer and David Straub for enlightening discussions and acknowledge the financial support from the Slovenian Research Agency (research core funding No. P1-0035 and J1-8137). }

\appendix

%
\section{Form factors}\label{app:FFs}
%

We employ the parameterization of the hadronic matrix elements in terms of the form factors as found in Ref.~\cite{Bailey:2015dka,Horgan:2013hoa}. First, for $ B \rightarrow K $ transitions we have
\begin{eqnarray}
\langle K (p) | \bar{s} \gamma^\mu b | B (k) \rangle &=& f_+ (q^2) \left( k^\mu + p^\mu - \frac{M^2_B - M^2_K}{q^2} q^\mu \right) + f_0 (q^2) \frac{M^2_B - M^2_K}{q^2} q^\mu \, , \\
\langle K (p) | i \bar{s} \sigma^{\mu \nu} b | B (k) \rangle &=& \frac{2 f_T (q^2)}{M_B + M_K} (k^\mu p^\nu + k^\nu p^\mu) \, , \\
\langle K (p) | \bar{s} b | B (k) \rangle &=& \frac{M^2_B - M^2_K}{m_b - m_s} f_0 (q^2) \,.
\end{eqnarray}
Similarly for $ B_{(s)} \rightarrow V $, where $ V = K^\ast, \phi $, we employ
\begin{eqnarray}
\langle V (p, \varepsilon) | \bar{s} \gamma^\mu b | {B_{(s)}} (k) \rangle &=& \frac{2 i V (q^2)}{M_{B_{(s)}} + M_V} \epsilon^{\mu \nu \rho \sigma} \varepsilon^\ast_\nu p_\rho k_\sigma \, , \\
\langle V (p, \varepsilon) | \bar{s} \gamma^\mu \gamma^5 b | {B_{(s)}} (k) \rangle &=& 2 M_V A_0 (q^2) \frac{\varepsilon^\ast \cdot q}{q^2} q^\mu + (M_{B_{(s)}} + M_V) A_1 (q^2) \left( \varepsilon^{\ast \mu} - \frac{\varepsilon^\ast \cdot q}{q^2} q^\mu \right) \nonumber\\
&& - A_2 (q^2) \frac{\varepsilon^\ast \cdot q}{M_{B_{(s)}} + M_V} \left[ (k + p)^\mu - \frac{M_{B_{(s)}}^2 - M_V^2}{q^2} q^\mu \right] \, , \\
q^\nu \langle V (p, \varepsilon) | \bar{s} \sigma_{\mu \nu} b | {B_{(s)}} (k) \rangle &=& 2 T_1 (q^2) \epsilon_{\mu \rho \tau \sigma} \varepsilon^{\ast \rho} k^\tau p^\sigma \, , \\
q^\nu \langle V (p, \varepsilon) | \bar{s} \sigma_{\mu \nu} \gamma^5 b | {B_{(s)}} (k) \rangle &=& i T_2 (q^2) [ (\varepsilon^\ast \cdot q)(k + p)_\mu - \varepsilon_\mu^\ast (M_{B_{(s)}}^2 - M_V^2) ] \\
&& + i T_3 (q^2) (\varepsilon^\ast \cdot q) \left[ \frac{q^2}{M_{B_{(s)}}^2 - M_V^2} (k+p)_\mu - q_\mu \right] \, , \nonumber
\end{eqnarray}
while $ \langle V (p, \varepsilon) | \bar{q} \gamma^5 b | {B_{(s)}} (k) \rangle $ can be determined from $ q_\mu \langle V (p, \varepsilon) | \bar{q} \gamma^\mu \gamma^5 b | {B_{(s)}} (k) \rangle $. In both cases we neglect the mass of the strange quark over the mass of the bottom quark.

 For the $ B \rightarrow M \ell^+ \ell^- $ form factors we employ determinations using lattice QCD. Fortunately, the  $ B \rightarrow M \tau^+ \tau^- $ phase-space is restricted to the high-$ q^2 $ (small $M$ recoil) region, where the current lattice QCD form factor calculations are directly applicable. For the  $ B \rightarrow K \ell^+ \ell^- $ process, we use the form factors given in Ref.~\cite{Bailey:2015dka}. They are parameterized in terms of $ \mathcal{B} = \{ b^+_{0,1,2}, b^0_{0,1,2}, b^T_{0,1,2} \} $ in the following way
\begin{equation}
f_{+,T} (q^2) = \frac{1}{P_{+,T} (q^2)} \sum^{K-1}_{m=0} b^{+,T}_m \left[ z^m - (-1)^{m-K} \frac{m}{K} z^K \right] \, , \quad
f_0 (q^2) = \frac{1}{P_0 (q^2)} \sum^{K-1}_{m=0} b^0_m z^m \, , \quad K = 3 \, ,
\end{equation}
where
\begin{equation}
z (q^2, t_0) = \frac{\sqrt{t_+ - q^2} - \sqrt{t_+ - t_0}}{\sqrt{t_+ - q^2} + \sqrt{t_+ - t_0}} \, , \quad
P_{+,0,T} (q^2) = 1 - q^2 / M^2_{+,0,T} \, ,
\end{equation}
with $ t_0 = (M_B + M_K) ( \sqrt{M_B} - \sqrt{M_K} )^2 $, $ t_+ = (M_B + M_K)^2 $, $ M_{+,T} = M_{B^*_s} = 5.4154 $~GeV, $ M_0 = M_{B^*_{s0}} = 5.711 $~GeV, $ M_B = 5.27958 $~GeV and $ M_K = 0.497614 $~GeV.

Lattice extractions of the form factors are also available for $ B \rightarrow K^* \ell^+ \ell^- $ and $ B_s \rightarrow \phi \ell^+ \ell^- $, and we use the results of Ref.~\cite{Horgan:2013hoa,Horgan:2015vla} (the differences with respect to \cite{Bouchard:2013mia}, such as a smaller lattice spacing, are commented therein). We note that in these calculations the $ K^* $ or $ \phi $ have been treated as stable particles (see e.g. \cite{Agadjanov:2016fbd} for a discussion related to this approximation). Here form factors are parameterized in terms of $ \mathcal{A} = \{ a^F_{0,1} \} $, $ F = V, A_{0,1,12}, T_{1,2,23} $, as follows
\begin{equation}
F (q^2) = \frac{a^F_0 + a^F_1 z (q^2, t_0)}{1 - q^2 / (M_{B_{(s)}} + \Delta M^F)^2} \, , \quad F = V, A_{0,1,12}, T_{1,2,23} \, ,
\end{equation}
with  $ t_0 = 12 $~GeV, $ t_+ = (M_{B_{(s)}} + M_V)^2 $, $ V = K^*, \phi $, $ M_{B_s} = 5.366 $~GeV, $ M_{K^*} = 0.892 $~GeV and $ M_\phi = 1.020 $~GeV, while the mass differences $ \Delta M^F $ are given in Ref.~\cite{Horgan:2013hoa}.  




\section{Full expressions}\label{app:expAsyms}

We now list the different expressions for the asymmetries and branching ratios discussed in the main text. In what follows, $ U = B_{(s)} $ and $ V = K^\ast, \phi $. First we have, for $ B \rightarrow K \tau \tau $:

\vspace{4mm}


\begin{dmath}
\Gamma (K) = \frac{\Phi}{3
   q^2} 4 \pi  \left(\, | f_A |^2
   \left(\delta  \lambda +24
   \, m_\tau^2 \, M_K^2
   q^2\right)+6 q^2
   (2 \, {\rm Re} [f_A^\ast \, f_P] \, m_\tau \xi
   +\, | f_P |^2 q^2)+\delta 
   \lambda  \, | f_V |^2\right) \, ,
\end{dmath}


\begin{dmath}
\mathcal{P}^-_L (K) \times \Gamma (K) = \frac{\Phi}{3
   \sqrt{q^2}}8 \pi  \, {\rm Re} [f_A^\ast \, f_V]
   \sqrt{\Delta } \lambda \, ,
\end{dmath}


\begin{dmath}
\mathcal{P}^-_L (K) + \mathcal{P}^+_L (K) = 0 \, ,
\end{dmath}


\begin{dmath}
\mathcal{P}^-_T (K) \times \Gamma (K) = \frac{\Phi}{\sqrt{q^2}} 8 \pi ^2 \sqrt{\lambda }
   (\, {\rm Re} [f_A^\ast \, f_V] \, m_\tau \xi
   +\, {\rm Re} [f_P^\ast \, f_V]
   q^2) \, ,
\end{dmath}


\begin{dmath}
\mathcal{P}^-_T (K) + \mathcal{P}^+_T (K) = 0 \, ,
\end{dmath}


\begin{dmath}
\mathcal{P}^-_N (K) \times \Gamma (K) = - \Phi 8 \pi ^2 \, {\rm Im} [f_A^\ast \, f_P]
   \sqrt{\Delta } \sqrt{\lambda } \, ,
\end{dmath}


\begin{dmath}
\Big( \mathcal{P}^-_N (K) - \mathcal{P}^+_N (K) \Big) \times \Gamma (K) = - \Phi 16 \pi ^2 \, {\rm Im} [f_A^\ast \, f_P] \sqrt{\Delta }
   \sqrt{\lambda } \, ,
\end{dmath}


\begin{multline}
\mathcal{P}_{LL} (K) \times \Gamma (K) = \frac{\Phi}{3 q^2} 4 \pi  \Big(\, | f_A |^2
   \left(10 \lambda 
   \, m_\tau^2+24 \, m_\tau^2
   \, M_K^2 q^2-\lambda 
   q^2\right) \\ +q^2 (12
   \, {\rm Re} [f_A^\ast \, f_P] \, m_\tau \xi +6
   \, | f_P |^2 q^2-\lambda 
   \, | f_V |^2)+2 \lambda 
   \, m_\tau^2
   \, | f_V |^2\Big) \, ,
\end{multline}


\begin{multline}
\mathcal{P}_{TT} (K) \times \Gamma (K) = -\frac{\Phi}{3 q^2} 16 \pi  \Big(\, | f_A |^2
   \left(10 \lambda 
   \, m_\tau^2+24 \, m_\tau^2
   \, M_K^2 q^2-\lambda 
   q^2\right) \\ +q^2 (12
   \, {\rm Re} [f_A^\ast \, f_P] \, m_\tau \xi +6
   \, | f_P |^2 q^2+\lambda 
   \, | f_V |^2)-2 \lambda 
   \, m_\tau^2
   \, | f_V |^2\Big) \, ,
\end{multline}


\begin{multline}
\mathcal{P}_{NN} (K) \times \Gamma (K) = -\frac{\Phi}{3
   q^2} 16 \pi  \Big(\, | f_A |^2
   \left(\delta  \lambda +24
   \, m_\tau^2 \, M_K^2
   q^2\right) \\ +6 q^2
   (2 \, {\rm Re} [f_A^\ast \, f_P] \, m_\tau \xi
   +\, | f_P |^2 q^2)+\delta 
   \lambda 
   (-\, | f_V |^2)\Big) \, ,
\end{multline}


\begin{dmath}
\mathcal{P}_{LT} (K) \times \Gamma (K) = -\frac{\Phi}{q^2} 4 \pi ^2 \sqrt{\Delta }
   \sqrt{\lambda } (\, | f_A |^2
   \, m_\tau \xi +\, {\rm Re} [f_A^\ast \, f_P]
   q^2) \, ,
\end{dmath}


\begin{dmath}
\mathcal{P}_{LT} (K) - \mathcal{P}_{TL} (K) = 0 \, ,
\end{dmath}


\begin{dmath}
\mathcal{P}_{LN} (K) \times \Gamma (K) = -\frac{\Phi}{\sqrt{q^2}} 4 \pi ^2 \sqrt{\lambda }
   (\, {\rm Im} [f_A^\ast \, f_V] \, m_\tau \xi
   +\, {\rm Im} [f_P^\ast \, f_V]
   q^2) \, ,
\end{dmath}


\begin{dmath}
\mathcal{P}_{LN} (K) - \mathcal{P}_{NL} (K) = 0 \, ,
\end{dmath}


\begin{dmath}
\mathcal{P}_{TN} (K) \times \Gamma (K) = \frac{\Phi}{3
   \sqrt{q^2}} 32 \pi  \, {\rm Im} [f_A^\ast \, f_V]
   \sqrt{\Delta } \lambda \, ,
\end{dmath}


\begin{dmath}
\mathcal{P}_{TN} (K) - \mathcal{P}_{NT} (K) = 0 \, ,
\end{dmath}


\begin{dmath}
\mathcal{A}_{FB} (K) = 0 \, .
\end{dmath}


\noindent
Now, for $ B \rightarrow V \tau \tau $:

\begin{multline}
\Gamma (V) = \frac{\Phi}{3
   \, M_V^2 q^2} 4 \pi  \Big(\lambda 
   \Big(2 \, | A |^2 \delta 
   \, M_V^2 q^2 \\ +2 \xi 
   \Big(\delta 
   (\, {\rm Re} [B^\ast \, C]+\, {\rm Re} [F^\ast \, G])+6
   \, {\rm Re} [G^\ast \, H] \, m_\tau^2
   q^2\Big)+\, | C |^2
   \delta  \lambda +2 \Delta 
   \, | E |^2 \, M_V^2
   q^2 \\ +12 \, {\rm Re} [F^\ast \, H]
   \, m_\tau^2 q^2+\delta 
   \, | G |^2 \lambda +24
   \, | G |^2 \, m_\tau^2
   \, M_V^2 q^2+6
   \, | H |^2 \, m_\tau^2
   (q^2)^2\Big) \\ +\, | B |^2
   \delta  \Big(\lambda +12
   \, M_V^2
   q^2\Big)+\, | F |^2
   \Big(\delta  \lambda +12
   \Delta  \, M_V^2
   q^2\Big)\Big) \, ,
\end{multline}


\begin{dmath}
\mathcal{P}^-_L (V) \times \Gamma (V) = \frac{\Phi}{3
   \, M_V^2 \sqrt{q^2}} 8 \pi  \sqrt{\Delta }
   \left(\lambda  \left(2
   \, {\rm Re} [A^\ast \, E] \, M_V^2
   q^2+\xi 
   (\, {\rm Re} [B^\ast \, G]+\, {\rm Re} [C^\ast \, F])+\, {\rm Re} [C^\ast \, G] \lambda
   \right)+\, {\rm Re} [B^\ast \, F]
   \left(\lambda +12 \, M_V^2
   q^2\right)\right) \, ,
\end{dmath}


\begin{dmath}
\mathcal{P}^-_L (V) + \mathcal{P}^+_L (V) = 0 \, ,
\end{dmath}


\begin{dmath}
\mathcal{P}^-_T (V) \times \Gamma (V) = \frac{\Phi}{\, M_V^2
   \sqrt{q^2}} 8 \pi ^2 \sqrt{\lambda }
   \, m_\tau \left(4 \, {\rm Re} [A^\ast \, B]
   \, M_V^2 q^2+\xi 
   (\, {\rm Re} [B^\ast \, F]+\, {\rm Re} [B^\ast \, H]
   q^2)+\, {\rm Re} [B^\ast \, G]
   \left(\lambda +4 \, M_V^2
   q^2\right)+\lambda 
   (\, {\rm Re} [C^\ast \, F]+\, {\rm Re} [C^\ast \, G] \xi
   +\, {\rm Re} [C^\ast \, H]
   q^2)\right) \, ,
\end{dmath}


\begin{dmath}
\Big( \mathcal{P}^-_T (V) + \mathcal{P}^+_T (V) \Big) \times \Gamma (V) = \Phi 64 \pi ^2 \, {\rm Re} [A^\ast \, B]
   \sqrt{\lambda } \, m_\tau
   \sqrt{q^2} \, ,
\end{dmath}


\begin{dmath}
\mathcal{P}^-_N (V) \times \Gamma (V) = -\frac{\Phi}{\, M_V^2} 8 \pi ^2 \sqrt{\Delta }
   \sqrt{\lambda } \, m_\tau
   \left(-2 \, M_V^2
   (\, {\rm Im} [A^\ast \, F]+\, {\rm Im} [B^\ast \, E]-2
   \, {\rm Im} [F^\ast \, G])+\, {\rm Im} [F^\ast \, H] \xi
   +\, {\rm Im} [G^\ast \, H] \lambda
   \right) \, ,
\end{dmath}


\begin{dmath}
\Big( \mathcal{P}^-_N (V) - \mathcal{P}^+_N (V) \Big) \times \Gamma (V) = -\frac{\Phi}{\, M_V^2} 16 \pi ^2 \sqrt{\Delta }
   \sqrt{\lambda } \, m_\tau
   \left(4 \, {\rm Im} [F^\ast \, G]
   \, M_V^2+\, {\rm Im} [F^\ast \, H] \xi
   +\, {\rm Im} [G^\ast \, H] \lambda
   \right) \, ,
\end{dmath}


\begin{dmath}
\mathcal{P}_{LL} (V) \times \Gamma (V) = \frac{\Phi}{3
   \, M_V^2 q^2} 4 \pi  \Big(\lambda 
   \Big(2 \, | A |^2 \, M_V^2
   q^2 \Big(2
   \, m_\tau^2-q^2\Big)+2
   \xi  \Big(2 \, m_\tau^2
   (\, {\rm Re} [B^\ast \, C]+5
   \, {\rm Re} [F^\ast \, G])-q^2
   \Big(\, {\rm Re} [B^\ast \, C]+\, {\rm Re} [F^\ast \, G]-
   6 \, {\rm Re} [G^\ast \, H]
   \, m_\tau^2\Big)\Big)+2
   \, | C |^2 \lambda 
   \, m_\tau^2-\, | C |^2 \lambda 
   q^2-2 \Delta  \, | E |^2
   \, M_V^2 q^2+12
   \, {\rm Re} [F^\ast \, H] \, m_\tau^2
   q^2 \\ +10 \, | G |^2 \lambda
    \, m_\tau^2+24 \, | G |^2
   \, m_\tau^2 \, M_V^2
   q^2-\, | G |^2 \lambda 
   q^2+6 \, | H |^2
   \, m_\tau^2
   (q^2)^2\Big) \\ +\, | B |^2
   \Big(2
   \, m_\tau^2-q^2\Big)
   \Big(\lambda +12 \, M_V^2
   q^2\Big)+\, | F |^2
   \lambda  \Big(10
   \, m_\tau^2-q^2\Big)-1
   2 \Delta  \, | F |^2
   \, M_V^2
   q^2\Big) \, ,
\end{dmath}


\begin{dmath}
\mathcal{P}_{TT} (V) \times \Gamma (V) = \frac{\Phi}{3
   \, M_V^2 q^2} 16 \pi  \Big(\, | B |^2
   \Big(2 \lambda 
   \, m_\tau^2+24 \, m_\tau^2
   \, M_V^2 q^2-\lambda 
   q^2\Big) \\ -\lambda 
   \Big(-4 \, | A |^2 \, m_\tau^2
   \, M_V^2
   q^2-\, | A |^2
   \, M_V^2 (q^2)^2 \\ +2 \xi 
   \Big(q^2
   \Big(\, {\rm Re} [B^\ast \, C]-\, {\rm Re} [F^\ast \, G]+
   6 \, {\rm Re} [G^\ast \, H]
   \, m_\tau^2\Big)-2
   \, m_\tau^2 (\, {\rm Re} [B^\ast \, C]-5
   \, {\rm Re} [F^\ast \, G])\Big) \\ -2
   \, | C |^2 \lambda 
   \, m_\tau^2+\, | C |^2 \lambda 
   q^2+\Delta  \, | E |^2
   \, M_V^2 q^2+10
   \, | F |^2
   \, m_\tau^2-\, | F |^2
   q^2+12 \, {\rm Re} [F^\ast \, H]
   \, m_\tau^2 q^2 \\ +10
   \, | G |^2 \lambda 
   \, m_\tau^2+24 \, | G |^2
   \, m_\tau^2 \, M_V^2
   q^2-\, | G |^2 \lambda 
   q^2+6 \, | H |^2
   \, m_\tau^2
   (q^2)^2\Big)\Big) \, ,
\end{dmath}


\begin{dmath}
\mathcal{P}_{NN} (V) \times \Gamma (V) = -\frac{\Phi}{3
   \, M_V^2 q^2} 16 \pi  \Big(\lambda 
   \Big(q^2 \Big(\Delta 
   \, M_V^2
   (\, | A |^2-\, | E |^2)+6
   \, | H |^2 \, m_\tau^2
   q^2\Big) \\ +2 \xi 
   \Big(-\, {\rm Re} [B^\ast \, C] \delta
   +\delta  \, {\rm Re} [F^\ast \, G]+6
   \, {\rm Re} [G^\ast \, H] \, m_\tau^2
   q^2\Big)+\delta 
   (\, | F |^2-\, | C |^2 \lambda
   ) \\ +12 \, {\rm Re} [F^\ast \, H] \, m_\tau^2
   q^2+\, | G |^2
   \Big(\delta  \lambda +24
   \, m_\tau^2 \, M_V^2
   q^2\Big)\Big)-\, | B |^2 \Big(\delta  \lambda +24
   \, m_\tau^2 \, M_V^2
   q^2\Big)\Big) \, ,
\end{dmath}


\begin{dmath}
\mathcal{P}_{LT} (V) \times \Gamma (V) = -\frac{\Phi}{\, M_V^2
   q^2} 4 \pi ^2 \sqrt{\Delta }
   \sqrt{\lambda } \, m_\tau
   \Big(-2 \, {\rm Re} [A^\ast \, F]
   \, M_V^2 q^2-2
   \, {\rm Re} [B^\ast \, E] \, M_V^2
   q^2+\, | F |^2 \xi +2
   \, {\rm Re} [F^\ast \, G] \lambda +4
   \, {\rm Re} [F^\ast \, G] \, M_V^2
   q^2+\, {\rm Re} [F^\ast \, H] \xi 
   q^2+\, | G |^2 \lambda 
   \, M_U^2-\, | G |^2 \lambda 
   \, M_V^2-\, | G |^2 \lambda 
   q^2+\, {\rm Re} [G^\ast \, H] \lambda 
   q^2\Big) \, ,
\end{dmath}


\begin{dmath}
\Big( \mathcal{P}_{LT} (V) - \mathcal{P}_{TL} (V) \Big) \times \Gamma (V) = \Phi 16 \pi ^2 \sqrt{\Delta }
   \sqrt{\lambda } \, m_\tau
   (\, {\rm Re} [A^\ast \, F]+\, {\rm Re} [B^\ast \, E]) \, ,
\end{dmath}


\begin{dmath}
\mathcal{P}_{LN} (V) = \frac{\Phi}{\, M_V^2
   \sqrt{q^2}} 4 \pi ^2 \sqrt{\lambda }
   \, m_\tau \left(\xi 
   (\, {\rm Im} [B^\ast \, F]+\, {\rm Im} [B^\ast \, H]
   q^2)+\, {\rm Im} [B^\ast \, G]
   \left(\lambda +4 \, M_V^2
   q^2\right)+\lambda 
   (\, {\rm Im} [C^\ast \, F]+\, {\rm Im} [C^\ast \, G] \xi
   +\, {\rm Im} [C^\ast \, H]
   q^2)\right) \, ,
\end{dmath}


\begin{dmath}
\mathcal{P}_{LN} (V) - \mathcal{P}_{NL} (V) = 0 \, ,
\end{dmath}


\begin{dmath}
\mathcal{P}_{TN} (V) \times \Gamma (V) = -\frac{\Phi}{3
   \, M_V^2 \sqrt{q^2}} 32 \pi  \sqrt{\Delta }
   \lambda  \left(-\, {\rm Im} [A^\ast \, E]
   \, M_V^2
   q^2+\, {\rm Im} [B^\ast \, F]+\xi 
   (\, {\rm Im} [B^\ast \, G]+\, {\rm Im} [C^\ast \, F])+\, {\rm Im} [C^\ast \, G] \lambda \right) \, ,
\end{dmath}


\begin{dmath}
\mathcal{P}_{TN} (V) - \mathcal{P}_{NT} (V) = 0 \, ,
\end{dmath}


\begin{dmath}
\mathcal{A}_{FB} (V) \times \Gamma (V) = \Phi 8 \pi  \sqrt{\Delta }
   \sqrt{\lambda }
   \sqrt{q^2}
   (\, {\rm Re} [A^\ast \, F]+\, {\rm Re} [B^\ast \, E]) \, .
\end{dmath}

\vspace{4mm}

Above, we employ the following notation:

\vspace{4mm}


\begin{dmath}
\Phi = \frac{\sqrt{\Delta / q^2} \sqrt{\lambda}}{64 \, M_U^3 \, \pi} \, ,
\end{dmath}

\begin{dmath}
f_A = 4 \, f_+ (q^2)
   (\, C_{10}+\, C'_{10}+\, \delta C_{10}) \, ,
\end{dmath}

\begin{dmath}
f_P = \frac{1}{q^2} 4 \, m_\tau
   (\, C_{10}+\, C'_{10}+\, \delta C_{10}) ((\, f_0 (q^2)-\, f_+ (q^2))
   (\, M_U-\, M_K)
   (\, M_U+\, M_K)+\, f_+ (q^2) q^2) \, ,
\end{dmath}

\begin{dmath}
f_V = \frac{1}{\, M_U+\, M_K} 4 (\, f_+ (q^2)
   (\, M_U+\, M_K)
   (\, C^{\rm eff}_{9}+\, C'_{9}+\, \delta C_{9})-2 \, f_T (q^2)
   \, m_b
   (\, C^{\rm eff}_{7} + \, \delta C_7 +\, C'_7)) \, ,
\end{dmath}

\begin{dmath}
A = \frac{4 V(q^2)
   (\, C^{\rm eff}_9+\, C'_9+
   \, \delta C_9)}{\, M_U+\, M_V}+\frac{8 \, m_b
   \, T_1(q^2)
   (\, C^{\rm eff}_7 + \delta C_7 +\, C'_7)}{q^2} \, ,
\end{dmath}

\begin{dmath}
B = -\frac{1}{q^2} 2 (\, M_U+\, M_V)
   (q^2
   \, A_1(q^2)
   (\, C^{\rm eff}_9-\, C'_9+
   \, \delta C_9)+2 \, m_b
   (\, M_U-\, M_V)
   \, T_2(q^2)
   (\, C^{\rm eff}_7 + \delta C_7 -\, C'_7)) \, ,
\end{dmath}

\begin{dmath}
C = \frac{2 \, A_2(q^2)
   (\, C^{\rm eff}_9-\, C'_9+
   \, \delta C_9)}{\, M_U+\, M_V}+4 \, m_b
   (\, C^{\rm eff}_7 + \delta C_7 -\, C'_7)
   \left(\frac{\, T_3(q^2)}{\, M_U^2-\, M_V^2}
   +\frac{\, T_2(q^2)}{q^2}\right) \, ,
\end{dmath}

\begin{dmath}
E = \frac{4 V(q^2)
   (\, C_{10}+\, C'_{10}+\, \delta C_{10})}{\, M_U+\, M_V} \, ,
\end{dmath}

\begin{dmath}
F = 2 \, A_1(q^2)
   (\, M_U+\, M_V)
   (-\, C_{10}+\, C'_{10}-\, \delta C_{10}) \, ,
\end{dmath}

\begin{dmath}
G = \frac{2 \, A_2(q^2)
   (\, C_{10}-\, C'_{10}+\, \delta C_{10})}{\, M_U+\, M_V} \, ,
\end{dmath}

\begin{dmath}
H = \frac{1}{q^2 (\, M_U+\, M_V)} 2
   (-\, C_{10}+\, C'_{10}-\, \delta C_{10}) (2 \, M_V
   (\, M_U+\, M_V)
   (\, A_0(q^2)-\, A_3(q^2))-q^2
   \, A_2(q^2)) \, ,
\end{dmath}

\begin{dmath}
\lambda = M_U^4 + M_V^4 + (q^2)^2 - 2 \, (M_U^2 \, M_V^2 + M_V^2 \, q^2 + q^2 \, M_U^2) \, ,
\end{dmath}

\begin{dmath}
\Delta = -4 \, m_\tau^2 + q^2 \, ,
\end{dmath}

\begin{dmath}
\xi = M_U^2 - M_V^2 - q^2 \, ,
\end{dmath}

\begin{dmath}
\delta = 2 \, m_\tau^2 + q^2 \, .
\end{dmath}


\newpage{\pagestyle{empty}\cleardoublepage}


\end{document}